\documentclass[reprint,amsmath,amssymb,aps,prl, superscriptaddress,floatfix]{revtex4-1}

\usepackage{hyperref}
\usepackage{graphicx}
\usepackage{mathtools}
\usepackage{multirow}
\usepackage{array}
\usepackage{tikz}

\newcolumntype{L}{>{$}l<{$}}

\DeclareMathOperator{\Tr}{Tr}

\begin{document}

\title{Piezoresistivity as a Fingerprint of Ferroaxial
Transitions}

\author{Ezra Day-Roberts}
\affiliation{School of Physics and Astronomy, University of Minnesota, Minneapolis,
MN 55455, USA}
\affiliation{Department of Chemical Engineering and Materials Science, University of Minnesota, MN 55455, USA}
\author{Rafael M. Fernandes}
\affiliation{School of Physics and Astronomy, University of Minnesota, Minneapolis,
MN 55455, USA}
\author{Turan Birol}
\affiliation{Department of Chemical Engineering and Materials Science, University of Minnesota, MN 55455, USA}
\date{\today}

\begin{abstract}

Recent progress in the understanding of the collective behavior of electrons and ions have revealed new types of ferroic orders beyond ferroelectricity and ferromagnetism, such as the ferroaxial state. The latter retains only rotational symmetry around a single axis and reflection symmetry with respect to a single mirror plane, both of which are set by an emergent electric toroidal dipole moment. Due to this unusual symmetry-breaking pattern, it has been challenging to directly measure the ferroaxial order parameter, despite the increasing attention this state has drawn. Here, we show that off-diagonal components of the piezoresistivity tensor (i.e., the linear change in resistivity under strain) transform the same way as the ferroaxial moments, providing a direct probe of such order parameters. We identify two new proper ferroaxial materials through a materials database search, and use first-principles calculations to evaluate the piezoconductivity of the double-perovskite CaSnF$_6$, revealing its connection to ferroaxial order and to octahedral rotation modes. 
\end{abstract}
\maketitle


The magnetic dipole moment is a prime example of an axial vector in physics. While it behaves like an ordinary (i.e., polar) vector under rotations, it is invariant under spatial inversion. Another type of axial vector that emerges in condensed matter systems is the electric toroidal dipole moment, also called the ferroaxial (or ``ferrorotational'') moment \cite{Hayami2018a, Hayami2018b, Litvin2008, Hlinka2014, Hlinka2016, Kusunose2024}. In contrast to the magnetic dipole moment, the electric toroidal moment is invariant under time reversal. Importantly, while a single electron does not have an intrinsic electric toroidal dipole moment, long-range ferroaxial order is enabled only by the collective behavior of electrons or the lattice. Indeed, several materials have been observed to undergo a so-called ferroaxial transition towards a state displaying a macroscopic ferroaxial moment -- analogous to the macroscopic magnetization that emerges below a ferromagnetic transition. Examples of proposed ferroaxial materials include LuFe$_2$O$_4$, RbCuCl$_3$, Mo$_3$Al$_2$C, and GdTe$_3$, which undergo electronically driven transitions \cite{Hlinka2016, Ikeda2015, Crama1981, Harada1982,Harada1983,Burch2022}, as well as CaMn$_7$O$_{12}$, Rb$_2$Cd$_2$(SO$_4$)$_3$, and NiTiO$_3$, which undergo transitions driven by their crystal structure \cite{Perks2012, Johnson2012, naliniand2002, Waskowska2010, Hayashida2020, Hayashida2021}. The `chirality density wave' proposed in URu$_2$Si$_2$ can also be considered as a collection of local axial moments \cite{Kung2015}.

Ferroaxial order leads to the spontaneous breaking of all rotational symmetries except those around the axis parallel to the electric toroidal dipole moment. As a result, it also breaks the reflection symmetry with respect to any mirror plane that includes the toroidal moment, while leaving the perpendicular mirror and inversion symmetries intact. Hence, the ferroaxial order is distinct from chirality, which breaks \textit{all} mirrors and inversion centers, and is equivalent to an electric toroidal monopole \cite{Inda2024, Kusunose2024}. This creates a major challenge in probing ferroaxial phase transitions, as there is no obvious conjugate field to the electric toroidal dipole moment \cite{Hlinka2016}, and only mangetotransport or higher order responses such as second-order magnetostriction are proposed as possibilities \cite{Kirikoshi2023, Hayami2023Unconv, Roy2022}. This is in contrast to other widely studied ``ferroic'' electronic states displaying ferroelectric, ferromagnetic, and nematic (or ferroelastic) orders, whose conjugate fields are electric fields, magnetic fields, and deviatoric strain respectively. Meanwhile, the lattice distortion pattern generated by ferroaxial order is often not associated with either infrared optical or acoustic phonon modes. While diffraction and optical probes may identify the presence of a nonzero ferroaxial moment \cite{Guo2023, Jin2020}, the required sensitivity can be challenging to be achieved, making it desirable to identify a macroscopic observable that is a direct fingerprint of ferroaxial moments.

In this regard, it is noteworthy that, in the case of other ferroic states, certain transport coefficients are linearly proportional to the corresponding ferroic order parameters. For instance, the anomalous Hall resistivity has the same symmetry properties as the magnetization \cite{Chen2020hall, Shao2020}, whereas the resistivity anisotropy is equivalent, on symmetry grounds, to the nematic order parameter \cite{Chu2010,Shapiro2015, Fernandes2012, Fradkin2010, Palmstrom2020Thesis}. Thus it is interesting to ask whether there is a transport coefficient that is symmetry-equivalent to the ferroaxial moment.

In this paper, by using group theory, phenomenology, and first-principles calculations, we show that the piezoresistivity can be used to directly probe the ferroaxial moments in a crystal (see schematics in Fig.~\ref{fig:experiment}). The piezoresistivity tensor $\Pi_{ijkl}$ measures the change in the resistivity $\rho_{ij}$ of a material that is linearly proportional to an applied strain $\varepsilon_{kl}$, i.e. 
\begin{equation}
\rho_{ij}=\Pi_{ijkl}\varepsilon_{kl} \label{eq:Pi}
\end{equation}
where the indices denote Cartesian coordinates and summation over repeated indices is implied. This quantity has the same symmetry properties as the linear elastoresistivity tensor defined in Ref. \cite{Shapiro2015}, $m_{ijkl} = \partial (\Delta \rho_{ij}/\rho)/\partial \varepsilon_{kl}$. It has been well-established that certain ``diagonal'' components of the piezoresistivity, such as $ \Pi_{xyxy}$ and  $(\Pi_{xxxx}- \Pi_{xxyy})$, are proportional to the nematic susceptibility \cite{Chu2012,Fernandes2012,Shapiro2015}. Here, we show that ``off-diagonal'' terms corresponding to changes in the longitudinal resistivity due to shear strain, such as $\Pi_{xxxy}$, or changes in the in-plane (out-of-plane) transverse resistivity due to an out-of-plane (in-plane) shear strain, such as $ \Pi_{xyyz}$, give the different components of the ferroaxial order parameter $\boldsymbol{\Psi}=(\Psi_x, \Psi_y, \Psi_z)$.

To support our symmetry analysis, we perform a materials search to identify candidate materials that display a ferroaxial transition. We identify the double-perovskite $\mathrm{CaSnF_6}$ and the langbeinite  $\mathrm{Rb_2Cd_2(SO_4)_3}$ as cubic systems that  display \textit{proper} ferroaxial order \cite{mayer1983struktur, naliniand2002}. This is to be contrasted with many of the materials listed before \cite{Kung2015, Ikeda2015}, as well as  $1T$-$\mathrm{TaS_2}$ \cite{Brouwer1980}, in which the ferroaxial order parameter is improper, i.e. secondary. Focusing on the case of $\mathrm{CaSnF_6}$, we perform first principles density functional theory calculations and explicitly calculate the piezoresistivity tensor via a Boltzmann transport approach \cite{pizzi2014boltzwann},  confirming our symmetry analysis and revealing the important contribution of octahedral rotations to the piezoresistivity of this compound. 

\begin{figure}
\centering
\includegraphics[width=0.90\linewidth]{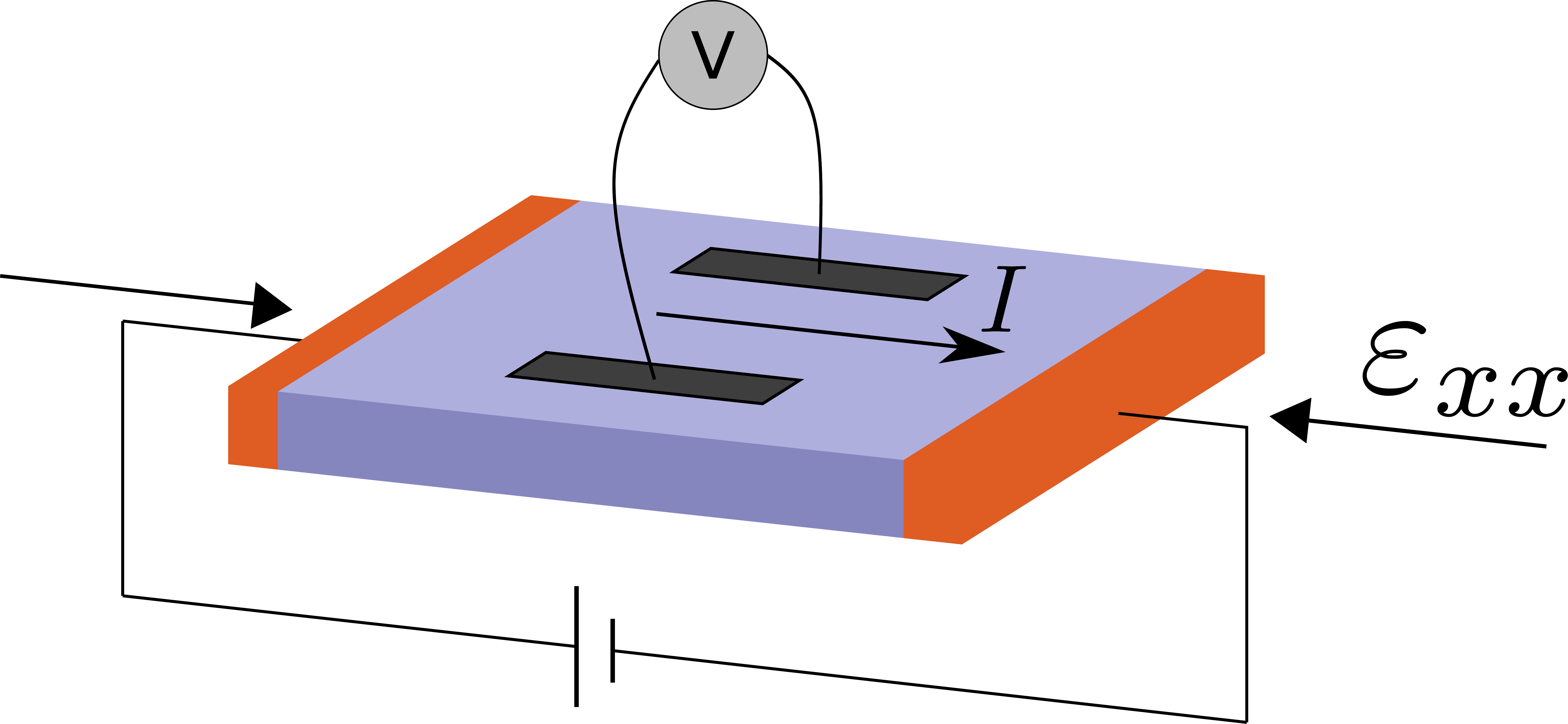}
\caption{Schematic of a DC piezoresistivity experiment.  The piezoresistivivity coefficient $\Pi_{xyxx}$ measures how uniaxial strain $\varepsilon_{xx}$ changes the dependence of the transverse voltage $V$ on the longitudinal current $I$ applied through the sample. }
\label{fig:experiment}
\end{figure}

Since piezoresistivity has been recently applied to extensively measure the nematic susceptibility of various correlated layered materials \cite{Chu2012,Riggs2015,Hosoi2016,Ishida2020,Eckberg2020}, it is illustrative to first present our results in this context to emphasize what is different in the case of ferroaxial order. Consider, for simplicity, the hypothetical case of an isotropic single layer. Electronic nematicity is formally described by a two-component vector $\boldsymbol{\eta} = (\eta_1, \eta_2)$, whose components correspond to some anisotropic response in the charge, orbital, or spin sector -- such as the uniform magnetic susceptibility, $\boldsymbol{\eta} \propto (\chi_{xx} - \chi_{yy}, 2 \chi_{xy})$, or $(p_x, p_y)$-orbital occupations, $\boldsymbol{\eta} \propto (n_{p_xp_x} - n_{p_yp_y}, 2 n_{p_x p_y})$ (see, for instance, Ref.~\cite{Fernandes2012}). In the disordered phase,  $\boldsymbol{\eta}=0$; however, the presence of symmetry-breaking in-plane strain necessarily induces a non-zero nematic order parameter, since $\boldsymbol{\varepsilon} = (\varepsilon_{xx} - \varepsilon_{yy}, 2 \varepsilon_{xy})$ also transforms as $\boldsymbol{\eta}$. In terms of the Landau free energy $F$ of the system, this effect arises from the bilinear term:

\begin{equation}
F_1= -\kappa_1 \left( \boldsymbol{\varepsilon} \cdot \boldsymbol{\eta} \right) \label{eq:F1}
\end{equation} 
where $\kappa_1$ is a coupling constant. Now, the electronic anisotropy encoded in $\boldsymbol{\eta}$ must be manifested in the transport properties via the anisotropic resistivity tensor $\boldsymbol{\rho}_\mathrm{ani} = (\rho_{xx} - \rho_{yy}, 2 \rho_{xy})$, since $\boldsymbol{\rho}_\mathrm{ani} \propto \boldsymbol{\eta}$ , as they have the same symmetry properties \cite{Chu2010,Fradkin2010}.  Therefore, Eq. (\ref{eq:F1})  gives the well-established relationship between the diagonal piezoresistivity coefficients and the nematic susceptibility, e.g. $\Pi_{xyxy} = \partial \rho_{xy} / \partial \varepsilon_{xy} \propto \chi_\mathrm{nem}$ \cite{Shapiro2015}.

Let us now discuss what changes in the case of ferroaxial order, which is described by the electric toroidal dipole moment $\boldsymbol{\Psi}$. In addition to Eq. \ref{eq:F1} , there is now a trilinear coupling between $\boldsymbol{\varepsilon}$, $\boldsymbol{\eta}$, and $\Psi_z$  of the form:

\begin{equation}
F_2= -\kappa_2 \, \Psi_z \left( \boldsymbol{\varepsilon} \times \boldsymbol{\eta} \right) \cdot \hat{\boldsymbol{z}} \label{eq:F2}
\end{equation}

 Using the fact that $\boldsymbol{\rho}_\mathrm{ani} \propto \boldsymbol{\eta}$, it is straightforward to conclude from Eq. (\ref{eq:F2})  that the off-diagonal piezoresistivity coefficients are proportional to the ferroaxial order parameter, namely, $\Pi_{xxxy} = \partial \rho_{xy} / \partial \varepsilon_{xx} = - \Pi_{xyxx} = - \partial \rho_{xy} / \partial \varepsilon_{xx} \propto \Psi_z$. Thus, while in the nematic case the order parameter is proportional to components of the resistivity tensor, thus enabling the nematic susptibility to be probed by the resistivity, in the ferroaxial case the order parameter is proportional to components of the piezoresistivity itself.

Group theory allows us to formalize this insight to any crystalline lattice. While the cases of layered tetragonal and hexagonal crystals are explained in the SM \cite{Supplement}, we here illustrate the procedure for the case of the cubic lattice, in which  the three components of the electric dipolar toroidal moment $\boldsymbol{\Psi}=(\Psi_x, \Psi_y, \Psi_z)$ transform as the same three-dimensional irreducible representation (irrep).

Being a rank-4 tensor, the piezoresistivity $\Pi_{ijkl}$ has 81 components. 
Under a point symmetry operation $R$,  represented by $3\times 3$ matrix $\alpha_R$ acting on the Cartesian components, it transforms as 
$
    \Pi_{ijkl} \rightarrow \alpha_{ii'} \alpha_{jj'} \alpha_{kk'} \alpha_{ll'} \Pi_{i'j'k'l'}  
$ \cite{Nye1985book}. 
The 81 components of $\Pi$  transform as an 81-dimensional reducible representation of the point group, which can be decomposed into irreps using the orthogonality theorem \cite{Jahn1949, dresselhaus}. 
This procedure is equivalent to building $81\times 81$ matrices representing each point group operation acting on the $81$-component vector representing the components of $\Pi$, then block diagonalizing these matrices and identifying each block with the irrep matrices listed in point group tables. 

Importantly, the rank-4 tensor $\Pi$ has the Jahn symbol [V$^2$][V$^2$], i.e., it is symmetric under the exchange of either the first two or the last two indices \cite{Jahn1949}\footnote{For a system with broken time reversal symmetry, the Jahn symbol becomes [V$^2$]$^*$[V$^2$].}. As a result, $\Pi$ can at most have 36 independent components, instead of 81. This allows us to use the Voigt notation and represent $\Pi$ by a $6\times 6$ matrix with row and column indices 1 to 6 corresponding to $\{xx, yy, zz, yz, zx, xy\}$, respectively. Performing the procedure outlined above for the cubic point group $m\bar{3}m$ ($\mathrm{O_h}$), we express the 36 components of $\Gamma^\Pi$  as irreps of the point group:
\begin{equation}
	\Gamma^\Pi = 3 A_{1g} \oplus A_{2g} \oplus 4 E_g \oplus 3 T_{1g} \oplus 5 T_{2g} .
\end{equation} 
Interestingly, every inversion-even irrep of the point group $m\bar{3}m$ appears in this expansion at least once. Note that no inversion-odd irrep appears, which excludes the possibility of probing chirality via piezoresistivity. Among the allowed irreps, there are $3$ independent non-zero components of $\Pi$ transforming as the 3 $A_{1g}$ irreps of the decomposition, and corresponding to $\tilde\Pi_{11} \equiv \Pi_{xxxx}$, $\Pi_{12} \equiv \Pi_{xxyy}$, and $\Pi_{44} \equiv \Pi_{yzyz}$, such that: 
\begin{align}
\Pi^{m\bar{3}m} &= \begin{pmatrix}
	\Pi_{11} & \Pi_{12} & \Pi_{12} & 0 & 0 & 0 \\
	\Pi_{12} & \Pi_{11} & \Pi_{12} & 0 & 0 & 0 \\
	\Pi_{12} & \Pi_{12} & \Pi_{11} & 0 & 0 & 0 \\
	0 & 0 & 0 & \Pi_{44} & 0 & 0 \\
	0 & 0 & 0 & 0 & \Pi_{44} & 0 \\
	0 & 0 & 0 & 0 & 0 & \Pi_{44}
\end{pmatrix}
\label{eq:Pi_prime}
\end{align}
The other components, which are zero in the cubic phase, correspond to symmetry-breaking electronic or structural order parameters. For example, $E_g$ and $T_{2g}$ represent nematic (or ferroelastic) order parameters \cite{Hecker2024}, whereas $A_{2g}$ corresponds to a $l=6$ (tetrahexacontapole) electric multipole, or equivalently, an electric toroidal octupole \cite{Hayami2018a}. Crucially for our purposes, the three-dimensional irrep $T_{1g}$, which appears three times in $\Gamma^\Pi$, corresponds to the ferroaxial order parameter $\boldsymbol{\Psi}$. Thus, its condensation leads to the onset of the corresponding non-zero piezoresistivity components. Since $\Pi$ includes each and every inversion-even irrep, different components of piezoresistivity can be used to probe any structural phase transition that does not break inversion. 

Focusing on ferroaxial order, we can build three axial vectors $\boldsymbol{\Psi}_n$, with $n=A, B, C$, from the components of $\Pi$:
\begin{equation}
	\boldsymbol{\Psi}_A=\frac{1}{2}\left(\Pi_{yyyz} - \Pi_{zzyz}, \Pi_{zzzx} - \Pi_{xxzx},  \Pi_{xxxy} - \Pi_{yyxy}\right)
\end{equation} 
\begin{equation}
	\boldsymbol{\Psi}_B=\frac{1}{2}\left(\Pi_{yzzz} - \Pi_{yzyy}, \Pi_{zxxx} - \Pi_{zxzz},  \Pi_{xyyy} - \Pi_{xyxx}\right)
\end{equation}
\begin{equation}
	\boldsymbol{\Psi}_C=\frac{1}{2}\left(\Pi_{xyzx} - \Pi_{zxxy}, \Pi_{yzxy} - \Pi_{xyyz},  \Pi_{zxyz} - \Pi_{yzzx}\right)
\end{equation} 
Therefore, the change in $\Pi$ across a ferroaxial (FA) transition can be expressed in terms of the components $\boldsymbol{\Psi}_n=(\Psi_{n1}, \Psi_{n2}, \Psi_{n3})$:

\begin{equation}
\Delta\Pi^\mathrm{FA}=
	\begin{pmatrix}
	0 & 0 & 0 & 0 & -\Psi_{A2} & \Psi_{A3} \\
	0 & 0 & 0 & \Psi_{A1} & 0 & -\Psi_{A3} \\
	0 & 0 & 0 & -\Psi_{A1} & \Psi_{A2} & 0 \\
	0 & -\Psi_{B1} & \Psi_{B1} & 0 & -\Psi_{C3} & \Psi_{C2} \\
	\Psi_{B2} & 0 & -\Psi_{B2} & \Psi_{C3} & 0 & -\Psi_{C1} \\
	-\Psi_{B3} & \Psi_{B3} & 0 & -\Psi_{C2} & \Psi_{C1} & 0
\end{pmatrix}
\label{eq:DeltaPi}
\end{equation}
Each $\boldsymbol{\Psi}$ has a different physical meaning: $\boldsymbol{\Psi}_A$ is the change in the longitudinal resistivity when a shear strain is applied;  $\boldsymbol{\Psi}_B$ is the change in the transverse resistivity when longitudinal strain is applied; and $\boldsymbol{\Psi}_C$ is the change in the in-plane (out-of-plane) transverse resistivity when an out-of-plane (in-plane) shear strain is applied. Since all three $\boldsymbol{\Psi}$'s transform as the same irrep, they must all be simultaneously zero or non-zero. The Landau theory of $\boldsymbol{\Psi}$, discussed in the SM, shows that in the absence of coupling with strain, the ferroaxial moment must point either along the cubic [111] body diagonals or the cubic [100] axes, resulting in eight or six domains, respectively \cite{Supplement}.

\begin{table}
\begin{ruledtabular}
\begin{tabular}{c|c|c|c}
Material & \shortstack{Low Temp \\ Space Group} & \shortstack{High Temp \\ Space Group} & $\Gamma$-point 
\\\hline
ZnTe \cite{Pellicer-Porres2001} & $P3_1$ & $F\overline{4}3m$ &   \\
CaSnF$_{6}$ \cite{mayer1983struktur} & $R\overline{3}$ & $Fm\overline{3}m$ & \checkmark  \\
Si \cite{Piltz1995} & $R\overline{3}$ & $Fd\overline{3}m$ &  \\
CsU$_{2}$O$_{6}$ \cite{vanEgmond1975} & $R\overline{3}$ & $Fd\overline{3}m$ &\\
Rb$_{2}$Cd$_{2}$(SO$_{4}$)$_{3}$ \cite{naliniand2002} & $P2_1$ & $P2_13$ & \checkmark \\
LiIO$_{3}$ \cite{Liang1989} & $P6_3$ & $P4_2/n$ &  \\
RbCuCl$_{3}$ \cite{Harada1982,Harada1983} & $C2$ & $I4/mcm$ & \\
\end{tabular}
\end{ruledtabular}
\caption{Ferroaxial material candidates that have reported structures both with and without an axial moment. All compounds except CaSnF$_{6}$ and Rb$_{2}$Cd$_{2}$(SO$_{4}$)$_{3}$ have a larger unit cell in the ferroaxial phase, as thus cannot be proper ferroaxial materials.}
\label{tab:good-transitions}
\end{table}

To proceed, we perform a materials search using the Materials Project Database \cite{Jain2013} to identify new nonmagnetic materials with experimentally observed transitions from a non-axial point group to an axial one. We list seven promising materials in Table \ref{tab:good-transitions}, of which two display proper ferroaxial order: the double-perovskite fluoride CaSnF$_6$  and the langbeinite  $\mathrm{Rb_2Cd_2(SO_4)_3}$. The other ones are improper ferroaxial materials, in that ferroaxial order is triggered by the condensation of a finite-momentum (zone-boundary) order parameter. 
While the phase transitions of these materials are well-established experimentally, they were not recognized to be ferroaxial before. 
In the remainder of the paper, we focus on CaSnF$_6$, shown in Fig.~\ref{fig:doublefluoride}. This compound is a member of the family of MM'X$_6$ cation-ordered perovskites with unoccupied  A-sites \cite{Evans2020}. Like most perovskites, it undergoes an anion octahedral rotation transition from a cubic high temperature $Fm\bar{3}m$ phase to a low temperature $R\bar{3}$ phase at 200 K \cite{gao2023new}. The rotation pattern is a$^-$a$^-$a$^-$ in Glazer notation, which would have led to the commonly observed $R\bar{3}c$ space group of single-perovskites if there was no cation order \cite{Lufaso2004, Glazer1972}. In CaSnF$_6$, however, the checkerboard ordering of cations doubles the unit cell, folding the zone boundary $R_5^-$ rotation mode onto the zone center $T_{1g}$ ferroaxial mode ($\Gamma_4^+$ irrep). This turns the octahedral rotations into a proper ferroaxial order parameter, similar to NiTiO$_3$ \cite{Hayashida2020}. 
\begin{figure}
\centering
\includegraphics[width=0.48\textwidth]{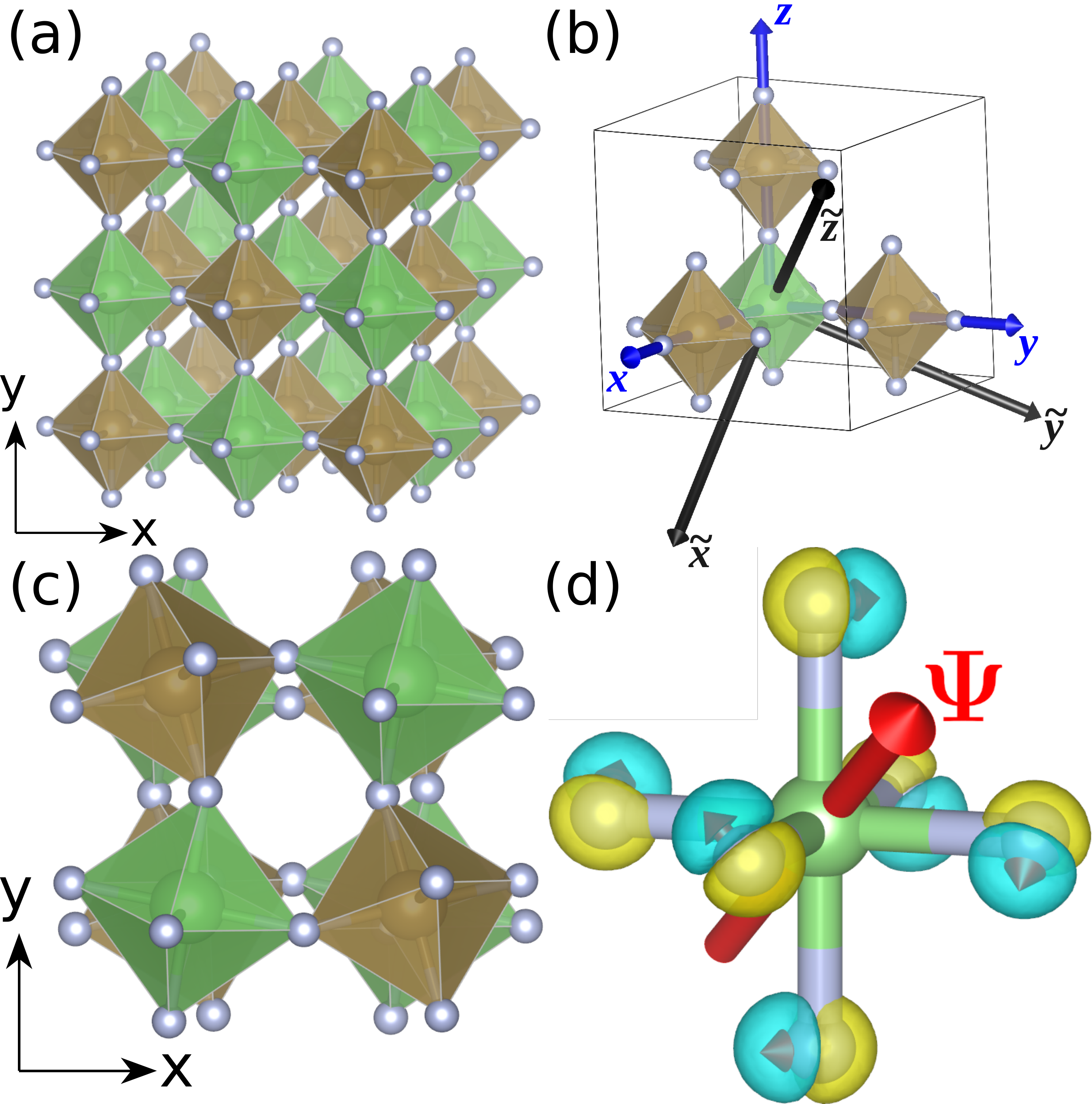}
\caption{(a) In the double perovskite structure of CaSnF$_6$, two different types of octahedra (green and brown) formed by the F atoms (grey spheres) enclose two types of metal atoms (Ca and Sn) in an alternating way. 
This staggered pattern of cations corresponds to a B-site R-point (i.e. $\vec{Q}=(\pi, \pi, \pi)$) cation order. (b) Cartesian coordinate axes for cubic (blue) and hexagonal (black, tilde) cells shown over the cubic structure. (c) The rhombohedral structure with octahedral rotations. (d) The atomic displacements (small arrows) due to the rotation of a single octahedron cause a change in electronic density (yellow/blue), resulting in an axial moment, $\Psi$, along the $[111]$ direction (large arrow).}
\label{fig:doublefluoride}
\end{figure}

To demonstrate that the octahedral rotations are indeed manifested in the piezoresistivity tensor of CaSnF$_6$, we calculate the latter via first-principles.
By convention, we use the Cartesian axes of the hexagonal unit cell of the low symmetry rhombohedral space group $R\bar{3}$~
instead of the pseudo-cubic axes (see Fig.~\ref{fig:doublefluoride} and SM for illustrations). We denote the hexagonal axes and quantities therein with tilde symbols for clarity. 
In this coordinate system, the ferroaxial moment points along $\hat{\tilde{z}}$, $\boldsymbol{\Psi} = \Psi \hat{\tilde{\boldsymbol{z}}}$, and the form of the piezoresistivity tensor in the symmetry-unbroken phase changes from Eq. (\ref{eq:Pi_prime}) to
\begin{align}
\tilde{\Pi}^{m\bar{3}m} &= \begin{pmatrix}
        \tilde{\Pi}_{11} & \tilde{\Pi}_{12} & \tilde{\Pi}_{13} & \tilde{\Pi}_{14} & 0 & 0 \\
        \tilde{\Pi}_{12} & \tilde{\Pi}_{11} & \tilde{\Pi}_{13} & -\tilde{\Pi}_{14} & 0 & 0 \\
        \tilde{\Pi}_{13} & \tilde{\Pi}_{13} & \tilde{\Pi}_{33} & 0 & 0 & 0 \\
        \tilde{\Pi}_{14} & -\tilde{\Pi}_{14} & 0 & \tilde{\Pi}_{44} & 0 & 0 \\
        0 & 0 & 0 & 0 & \tilde{\Pi}_{44} & \tilde{\Pi}_{14} \\
        0 & 0 & 0 & 0 & \tilde{\Pi}_{14} & \tilde{\Pi}_{66}
\end{pmatrix}
\label{eq:Pi_HS}
\end{align}
The 7 nonzero components $\tilde{\Pi}_{ij}$ are not independent, and can be expressed in terms of the three independent components $\Pi_{11}$, $\Pi_{12}$, and $\Pi_{44}$ of the piezoresistivity in the pseudocubic coordinate system of Eq. (\ref{eq:Pi_prime}), see SM \cite{Supplement}. Conversely, in the hexagonal coordinate system, the form of $\Delta\tilde{\Pi}^\mathrm{FA}$, i.e. the change in the piezoresistivity tensor due to ferroaxial order, is also different from Eq. (\ref{eq:DeltaPi}). 
While the full expression for $\Delta\tilde{\Pi}^\mathrm{FA}$ is given in the SM, we focus here on the elements $\Delta\tilde{\Pi}^\mathrm{FA}_{15} = -2(\Psi_A - 2\Psi_B + 2\Psi_C)$ and $\Delta\tilde{\Pi}^\mathrm{FA}_{16} = -2(\Psi_A + \Psi_B + 2\Psi_C)$, which can be directly read off from the piezoresistivity tensor calculated in the symmetry-broken phase, $\Delta\tilde{\Pi}^\mathrm{FA}_{15} = \tilde{\Pi}_{15}^{R\bar{3}}$ and $\Delta\tilde{\Pi}^\mathrm{FA}_{16} = \tilde{\Pi}_{16}^{R\bar{3}}$. This is not the case, however, for $\Delta\tilde{\Pi}^\mathrm{FA}_{14}$, since $\Delta\tilde{\Pi}^\mathrm{FA}_{14} \neq \tilde{\Pi}_{14}^{R\bar{3}}$ due to the fact that $\tilde{\Pi}_{14}^{m\bar{3}m} \neq 0$ in Eq. (\ref{eq:Pi_HS}).   

Because CaSnF$_{6}$ is a wide band-gap insulator with negligible conductivity, we consider hole doping, which can be achieved by cation vacancies or gating \cite{Goldman2014, Leighton2022}. We compute the conductivity tensor elements $\tilde{\sigma}_{ij}$ in the hexagonal coordinate system via the Boltzmann transport approach \cite{pizzi2014updated} as a function of uniaxial strain $\tilde{\varepsilon}_{xx}$ by building Wannier-based tight binding models at different strain values \cite{wannier90,wannierTheory,Supplement}.
This procedure gives the piezoconductivity tensor which, crucially, has the same symmetry properties as piezoresistivity. 
\begin{figure}
    \centering
    \includegraphics[width=0.48\textwidth]{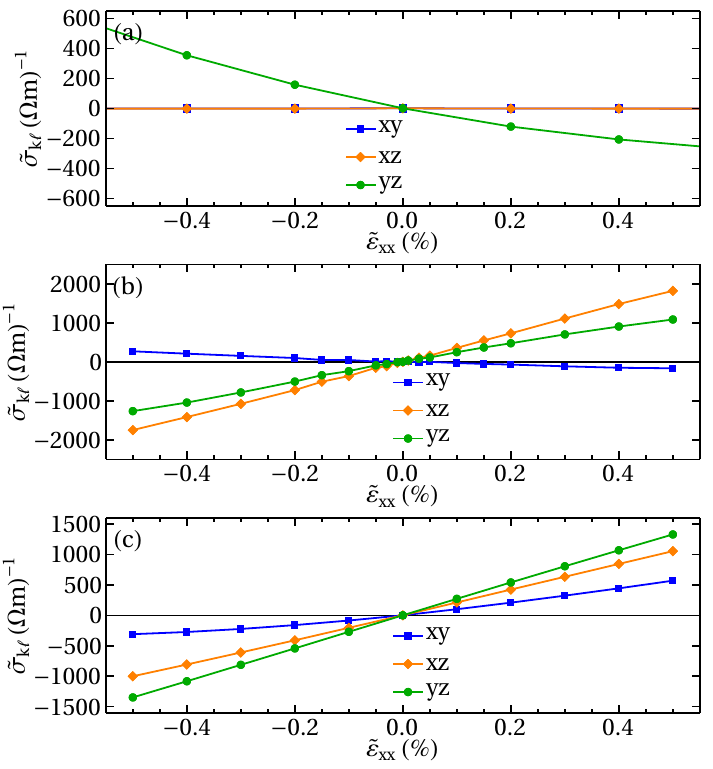}
    \caption{Off-diagonal components of the conductivity under uniaxial strain (a) in the cubic non-axial phase and (b) in the rhombohedral ferroaxial phase. (c) Same as in (b), but calculated without relaxing the internal coordinates of the atoms after strain is applied. }
    \label{fig:cond-plots}
\end{figure}

In Fig. \ref{fig:cond-plots}(a), we show the three transverse conductivities ($\tilde{\sigma}_{xy}$, $\tilde{\sigma}_{xz}$, and $\tilde{\sigma}_{yz}$) as a function of $\tilde{\varepsilon}_{xx}$ strain for a doping level of $0.05$ holes per metal atom (corresponding to $7\times 10^{20}$~cm$^{-3}$ carriers) in the cubic (non-ferroaxial) phase. In the absence of strain, as expected, the transverse conductivities vanish as required by symmetry. In the presence of strain, we find that $\tilde{\sigma}_{xz}$ and $\tilde{\sigma}_{xy}$  remain zero, i.e. $\partial\tilde{\sigma}_{xz}/\partial \tilde{\varepsilon}_{xx} = \partial\tilde{\sigma}_{xy}/\partial \tilde{\varepsilon}_{xx} = 0$, which is in agreement with the fact that $\tilde{\Pi}_{15}^{m\bar{3}m}=\tilde{\Pi}_{16}^{m\bar{3}m}=0$ in Eq. (\ref{eq:Pi_HS}). Conversely, $\tilde{\sigma}_{yz}$ depends linearly on strain, consistent with $\tilde{\Pi}_{14}^{m\bar{3}m}\neq0$. 
Moving on to the rhombohedral (ferroaxial) phase, shown in Fig.~\ref{fig:cond-plots}(b), we see a remarkable change in the behavior of  $\tilde{\sigma}_{xz}$ and $\tilde{\sigma}_{xy}$, which now display a linear dependence on strain $\tilde{\varepsilon}_{xx}$. This implies that $\Delta \tilde{\Pi}_{15}^{R\bar{3}} = \Delta\tilde{\Pi}^\mathrm{FA}_{15}\neq 0$ and $\tilde{\Pi}_{16}^{R\bar{3}} = \Delta\tilde{\Pi}^\mathrm{FA}_{16}\neq 0$, thus demonstrating that the off-diagonal piezoresistivity (piezoconductivity) are only non-zero inside the ferroaxial phase.

From Fig.~\ref{fig:cond-plots}(b), we extract the slopes (the off-diagonal piezoconductivities) as $\partial\tilde{\sigma}_{xy}/\partial \tilde{\varepsilon}_{xx} = -4.3 \times 10^4$ ($\Omega \textrm{m}$)$^{-1}$, $\partial\tilde{\sigma}_{xz}/\partial \tilde{\varepsilon}_{xx} = 3.6 \times 10^5$ ($\Omega \textrm{m}$)$^{-1}$, and $\partial\tilde{\sigma}_{yz}/\partial\tilde{\varepsilon}_{xx} = 2.4 \times 10^5$ ($\Omega \textrm{m}$)$^{-1}$. Because atomic positions were relaxed when different strains were imposed, these piezoconductivity values contain both the direct effect of strain (clamped-ion effects) and the effects mediated through changes in the atomic positions (including octahedral rotation angles) under strain. To disentangle these effects, we show in Fig. \ref{fig:cond-plots}(c) the off-diagonal piezoconductivity obtained after not relaxing the internal positions of atoms, which leaves only the direct effect of strain. Some of the slopes in this case are very different, $\partial\tilde{\sigma}_{xy}/\partial \tilde{\varepsilon}_{xx} = 8.9 \times 10^4$ ($\Omega \textrm{m}$)$^{-1}$, $\partial\tilde{\sigma}_{xz}/\partial \tilde{\varepsilon}_{xx}=2.1 \times 10^5$ ($\Omega \textrm{m}$)$^{-1}$, and $\partial\tilde{\sigma}_{yz}/\partial\tilde{\varepsilon}_{xx} = 2.7 \times 10^5$ ($\Omega \textrm{m}$)$^{-1}$, signaling the importance of the coupling between octahedral rotations, strain, and electronic structure in determining the magnitude of the piezoresistive response. This is likely a result of the electronic hopping parameters being more sensitive to the F-(Ca,Sn)-F bond angles than strain. While our calculations were performed in the metallic phase, it is in principle possible to measure the piezoresistivity of insulating materials, provided that the resistivity near the ferroaxial transition is not too large compared to the changes caused by strain. In this regard, a promising candidate ferroaxial material is 1$T$-TaS$_2$ \cite{Li2012,luo2021ultrafast,liu2023electrical}, which we discuss in detail in the Supplementary Material.

In summary, by considering the application of irrep projection operators on response tensors, we demonstrated that the off-diagonal components of the piezoresistivity can be used to directly measure the ferroaxial order parameter. 
By performing a materials database search, and then computing the piezoconductivity from first-principles, we discovered that the transitions observed in CaSnF$_6$ and Rb$_{2}$Cd$_{2}$(SO$_{4}$)$_{3}$ are proper ferroaxial transitions, in contrast to most other materials that exhibit only an improper ferroaxial transition. 

More broadly, our work underlines the capabilities of piezoresistivity as a powerful experimental probe to detect non-inversion-symmetry-breaking transitions in materials. This further extends the functionality of this response tensor, whose diagonal components have been widely employed over the past decade to obtain invaluable information about the nematic susceptibility of various materials.

\acknowledgments
We thank I. Fisher and Q. Jiang for fruitful discussions. E.D.R. and T.B. were supported by the National Science Foundation through the University of Minnesota MRSEC under Award Number DMR-2011401. R.M.F. was supported by the Air Force Office of Scientific Research under Grant No. FA9550-21-1-0423.

\pagebreak
\widetext
\setcounter{equation}{0}
\setcounter{figure}{0}
\setcounter{table}{0}
\setcounter{page}{1}
\makeatletter
\renewcommand{\theequation}{S\arabic{equation}}
\renewcommand{\thefigure}{S\arabic{figure}}
\renewcommand{\bibnumfmt}[1]{[S#1]}
\renewcommand{\citenumfont}[1]{S#1}

\pagebreak

\begin{center}
\textbf{Supplemental Information for \\ ``Piezoresistivity as an Order Parameter for Ferroaxial Transitions''}
\end{center}

\author{Ezra Day-Roberts}
\affiliation{School of Physics and Astronomy, University of Minnesota, Minneapolis,
MN 55455, USA}
\affiliation{Department of Chemical Engineering and Materials Science, University of Minnesota, MN 55455, USA}
\author{Rafael M. Fernandes}
\affiliation{School of Physics and Astronomy, University of Minnesota, Minneapolis,
MN 55455, USA}
\author{Turan Birol}
\affiliation{Department of Chemical Engineering and Materials Science, University of Minnesota, MN 55455, USA}
\date{\today}

\maketitle

\section{Details of DFT calculations}

Our calculations are implemented in the VASP software package \cite{VASP} using projected augmented wave methods \cite{VASP-PAW}. The exchange correlation effects were captured in the generalized gradient approximation (GGA) using the Perdew-Burke-Ernzerhof (PBE) functional finetuned for solids \cite{PBEsol}. A $\Gamma$ centered 5 x 5 x 5 $k$-point mesh was used in the one formula unit rhombohedral cell for all structures and calculations. A force cutoff of 1 meV/\AA ~was used for relaxations. The transport properties were calculated in the semiclassical Boltzmann transport theory via the Boltwann package \cite{pizzi2014boltzwann} using a tight-binding model for the valence band manifold fit by the Wannier90 code \cite{wannier90}. For each phase we calculate a separate Wannier model for each value of strain. The transport calculations where done on a 200 x 200 x 200 $k$ point mesh with a relaxation time of 10 fs. This relaxation time is a purely phenomenological value and the exact values will depend on the extrinsic details of the system. However this value can usually give plausible values.

In order to avoid issues with relaxation of charged cells in DFT calculations \cite{Brubeval201}, as well as the prediction of elastic and mechanical properties such as the Poisson ratios, the lattice vectors that define the unit cell are not relaxed in strain calculations. (This includes the calculations where the positions of the atoms within the unit cell are relaxed.) As a result, when we consider a $\varepsilon_{xx}$ strain applied on the system, our results do not include the $\varepsilon_{yy}$ and $\varepsilon_{zz}$ components. The effect of these components, which would be present in the experimental measurements, can be taken into account by considering the Poisson ratio of the system. However, they do not make a difference in the symmetry properties of response, and hence do not affect the symmetry discussions presented in the main text.

\section{Tensor Decomposition}
The general form for the projection operator onto a representation $\Gamma_n$ is 
\begin{equation}
	\hat P^{\Gamma_n} = \frac{\ell_n}{h} \sum_R \chi^{(\Gamma_n)}(R)^* \hat R
\end{equation}
where $R$ ranges over all symmetry operations of a specific group, $\chi^{(\Gamma)}$ is the character function of the representation $\Gamma$, $\ell_n$ is the dimensionality of the representation, and $h$ is the overall number of symmetry operations. For a derivation, see e. g. section 4.5 in Ref.~\cite{dresselhaus}. Since we are not interested in normalization we will drop the pre-factor. The usual issue with this formula is that we do not know how a symmetry operator acts on an arbitrary object. However as we are decomposing tensors the transformation is defined by the basic tensor property. Taking the cartesian transformation matrix of an operator to be $R_{ab}$, we can write the projection of a tensor $T$ of dimension $k$ onto representation $\Gamma$ as
\begin{equation}
	\hat P^{\Gamma} T_{i_1 i_2\dots i_k} \propto \sum_R \chi^{(\Gamma)}(R)^* R_{i_1i'_1} R_{i_2i'_2} \cdots R_{i_ki'_k} T_{i'_1 i'_2 \dots i'_k}
\end{equation}
By making our starting tensor arbitrary (symbolic) we obtain the general form of the contribution of that irrep.

By counting the number of independent variables remaining after projection and dividing by the dimensionality of the irrep we can also determine the multiplicity of each irrep in the tensor. This includes the case where the tensor does not contain any combination of components that transform as that irrep, in which case the projection will yield a zero tensor.

There is also a more cumbersome, but somewhat more direct way to obtain the multiplicity. One can explicitly construct the regular representation of the tensor by flattening the tensor into a large vector; for a rank-4 tensor with no symmetry this will be $3^4=81$ dimensional. The transformation properties of this can be deduced from the transformation properties of the tensor. Denoting the transformation matrix in the regular representation associated with operation $R$ as $D^R$ we have
\begin{equation}
    D^R_{(i_1,i_2,\hdots,i_d),(j_1,j_2,\hdots,j_d)} \equiv R_{i_1,j_1} R_{i_2,j_2} \cdots R_{i_d,j_d}
\end{equation}
Then a different form of the Orthogonality Theorem can be used to count the multiplicity for each irrep
\begin{equation}
    n^\Gamma = \frac{1}{h} \sum_{R} \chi^\Gamma(R) \Tr D^R
\end{equation}

\subsection{Example}

We can illustrate a simple use of this projection scheme with the example of a symmetric rank-2 tensor (such as conductivity) in point group $m$. The most general rank-2 tensor is
\begin{equation}
    \sigma = \left(
\begin{array}{ccc}
 \sigma _{11} & \sigma _{12} & \sigma _{13} \\
 \sigma _{12} & \sigma _{22} & \sigma _{23} \\
 \sigma _{13} & \sigma _{23} & \sigma _{33} \\
\end{array}
\right). 
\end{equation}
There are only two symmetry operations in this point group: the identity $E$ and the mirror $m$, which is chosen to be on the $xz$ plane following the crystallographic convention, and two irreps, $A'$ and $A''$, with characters
\begin{center}
\begin{tabular}{|c|c|c|}
\hline
     &  $\chi(E)$ & $\chi(m)$\\
     \hline
 $A'$    & 1&1\\
 $A''$  & 1 & -1\\
 \hline
\end{tabular}
\end{center}
The projection onto $A'$ is
\begin{align}
    \sigma^{A'} &= \chi^{A'}(E) \hat E \sigma + \chi^{A'}(m) \hat m\sigma \\
    &= (1) 
\begin{pmatrix}
 1 & 0 & 0 \\
 0 & 1 & 0 \\
 0 & 0 & 1 \\
\end{pmatrix}
\begin{pmatrix}
 \sigma _{11} & \sigma _{12} & \sigma _{13} \\
 \sigma _{12} & \sigma _{22} & \sigma _{23} \\
 \sigma _{13} & \sigma _{23} & \sigma _{33} \\
\end{pmatrix}
\begin{pmatrix}
 1 & 0 & 0 \\
 0 & 1 & 0 \\
 0 & 0 & 1 \\
\end{pmatrix}
+ (1) \begin{pmatrix}
     1 & 0 & 0 \\
 0 & -1 & 0 \\
 0 & 0 & 1 \\
\end{pmatrix}
\begin{pmatrix}
 \sigma _{11} & \sigma _{12} & \sigma _{13} \\
 \sigma _{12} & \sigma _{22} & \sigma _{23} \\
 \sigma _{13} & \sigma _{23} & \sigma _{33} \\
\end{pmatrix}
\begin{pmatrix}
     1 & 0 & 0 \\
 0 & -1 & 0 \\
 0 & 0 & 1 \\
\end{pmatrix} \\
&= \begin{pmatrix}
     \sigma _{11} & 0 & \sigma _{13} \\
 0 & \sigma _{22} & 0 \\
 \sigma _{13} & 0 & \sigma _{33} \\
\end{pmatrix} 
\end{align}

The projection operator onto $A''$ is different only in the character of $m$ in this irrep
\begin{align}
    \sigma^{A''} &= \chi^{A''}(E) \hat E \sigma + \chi^{A''}(m) \hat m\sigma \\
    &= (1) 
\begin{pmatrix}
 1 & 0 & 0 \\
 0 & 1 & 0 \\
 0 & 0 & 1 \\
\end{pmatrix}
\begin{pmatrix}
 \sigma _{11} & \sigma _{12} & \sigma _{13} \\
 \sigma _{12} & \sigma _{22} & \sigma _{23} \\
 \sigma _{13} & \sigma _{23} & \sigma _{33} \\
\end{pmatrix}
\begin{pmatrix}
 1 & 0 & 0 \\
 0 & 1 & 0 \\
 0 & 0 & 1 \\
\end{pmatrix}
+ (-1) \begin{pmatrix}
     1 & 0 & 0 \\
 0 & -1 & 0 \\
 0 & 0 & 1 \\
\end{pmatrix}
\begin{pmatrix}
 \sigma _{11} & \sigma _{12} & \sigma _{13} \\
 \sigma _{12} & \sigma _{22} & \sigma _{23} \\
 \sigma _{13} & \sigma _{23} & \sigma _{33} \\
\end{pmatrix}
\begin{pmatrix}
     1 & 0 & 0 \\
 0 & -1 & 0 \\
 0 & 0 & 1 \\
\end{pmatrix} \\
&= \begin{pmatrix}
  0 & \sigma _{12} & 0 \\
 \sigma _{12} & 0 & \sigma _{23} \\
 0 & \sigma _{23} & 0 \\
\end{pmatrix} 
\end{align}

As both are one dimensional irreps we can conclude that in this point group a rank two symmetric tensor has decomposition $4A' \oplus 2A''$. Since $A'$ is the fully symmetric irrep, everything that transforms as $A'$ is in principle nonzero when no symmetry is broken. The form of $\sigma^{A'}$ coincides with the form of a symmetric rank-2 tensor in the monoclinic point group $m$, as tabulated in, for example, Ref.~\cite{Nye1985book}. 

\section{Piezoresistivity in other point groups}
In $6/mmm$ the piezoresistivity has the decomposition 
\begin{equation}
    \Gamma^\Pi = 6 A_{1g} \oplus 2 A_{2g} \oplus 2 B_{1g} \oplus 2 B_{2g} \oplus 6 E_{1g} \oplus 6 E_{2g}
\end{equation}
In this group, the ferroaxial moment transforms as the sum of two separate irreps, $A_{2g} \oplus E_{1g}$, corresponding to the $c$-axis component and the in-plane components, respectively. Other quantities, such as deviatoric strain and electric quadrupolar order (nematic order), also transform as $E_{1g}$, which means that in-plane ferroaxial moments leaves the same signatures in the piezoresistivity as electronic nematic order. As a result, we only list the two $A_{2g}$ components here,
\begin{equation}
    \Delta\Pi^{A_{2g}} = \begin{pmatrix}
        0 & 0 & 0 & 0 & 0 & \Psi_{A3} \\
        0 & 0 & 0 & 0 & 0 & -\Psi_{A3} \\
        0 & 0 & 0 & 0 & 0 & 0 \\
        0 & 0 & 0 & 0 & \Psi_{B3} & 0 \\
        0 & 0 & 0 & -\Psi_{B3} & 0 & 0 \\
        \Psi_{A3} & -\Psi_{A3} & 0 & 0 & 0 & 0 
    \end{pmatrix}
\end{equation}

The form of the piezoresistivity tensor and its components for $4/mmm$ can be deduced by the subduction relations between $m\overline{3}m$ and $4/mmm$ but we nevertheless list them explicitly \cite{Aroyo2006}. In this group,  piezoresistivity has the decomposition
\begin{equation}
    \Gamma^\Pi = 7 A_{1g} \oplus 3 A_{2g} \oplus 5 B_{1g} \oplus 5 B_{2g} \oplus 16 E_g
\end{equation}
and the ferroaxial moment splits into $A_{2g} \oplus E_g$ for the out of plane and in plane components. This means the $A_{2g}$ components are merely the $z$ components from the $m\overline{3}m$ case,
\begin{equation}
    \Delta\Pi^{A_{2g}} = \begin{pmatrix}
        0 & 0 & 0 & 0 & 0 & \Psi_{A3} \\
        0 & 0 & 0 & 0 & 0 & -\Psi_{A3} \\
        0 & 0 & 0 & 0 & 0 & 0 \\
        0 & 0 & 0 & 0 & \Psi_{C3} & 0 \\
        0 & 0 & 0 & -\Psi_{C3} & 0 & 0 \\
        \Psi_{B3} & -\Psi_{B3} & 0 & 0 & 0 & 0 
    \end{pmatrix}
\end{equation}

\section{Application to 1$T$-TaS$_2$}
The layered transition-metal dichalcogenide 1$T$-TaS$_2$ undergoes a series of near-commensurate and commensurate charge density wave (CDW) transitionsthat reduce its space group from $P\overline{3}m1$ in its high temperature phase to $P\overline{3}$ in the low-temperature phase \cite{wilson1975charge}. This induces an out-of-plane ferroaxial moment that has been studied recently and reported to emerge at the near-commensurate CDW transition of approximately $350$ K \cite{liu2023electrical,luo2021ultrafast}. However, it is important to emphasize that the ferroaxial component is not the primary order parameter. The ``star of david'' distortion changes the unit cell to a $\sqrt{13}\times\sqrt{13}$ supercell coming from a low symmetry CDW wave-vector distortion. This reduces the symmetry to $P\overline{3}$ which then allows the ferroaxial irrep $A_{2g}$ as a secondary order parameter. Although  1$T$-TaS$_2$ is an insulator, the resistivity values observed near the $350$ K transition temperature are relatively small, of the order of $1\,\mathrm{m}\Omega.\mathrm{cm}$ , which may enable measurements of the piezoresistivity \cite{Li2012}.  Interestingly Fe-doping further reduces these resistivity values.
\section{Unit cell transformations for double-perovskites}

Here we discuss the coordinate systems change from the cubic unit cell of CaSnF$_6$ to the hexagonal one. We start from the cubic unit cell (black arrows) shown in Figure 3 with lattice vectors $(\boldsymbol{a}, \boldsymbol{b}, \boldsymbol{c})$. The matrix that transforms  these basis vectors to the hexagonal cell basis vectors $(\boldsymbol{\tilde{a}}, \boldsymbol{\tilde{b}}, \boldsymbol{\tilde{c}})$ is given by
\begin{equation}
\begin{pmatrix}
    \boldsymbol{\tilde{a}} \\ \boldsymbol{\tilde{b}} \\ \boldsymbol{\tilde{c}}
\end{pmatrix}
=
\begin{pmatrix}
 -\frac{1}{2} & \frac{1}{2} & 0 \\
 \frac{1}{2} & 0 & -\frac{1}{2} \\
 1 & 1 & 1 \\
\end{pmatrix}
\begin{pmatrix}
    \boldsymbol{a} \\ \boldsymbol{b} \\ \boldsymbol{c}
\end{pmatrix}
\end{equation}

The orthogonal axes corresponding to this setting have the $x$-axis along the first direction, the $z$-axis along the third and the $y$-axis chosen to be orthogonal to them. Thus, the corresponding relationship between the Cartesian coordinate systems of the cubic primitive cell (blue, primed arrows in Supplementary Fig. 1) and those of the hexagonal cell (black, unprimed arrows) is
\begin{equation}
\begin{pmatrix}
\boldsymbol{\tilde{x}} \\ \boldsymbol{\tilde{y}} \\ \boldsymbol{\tilde{z}}\end{pmatrix}
=
	\begin{pmatrix}
	\frac{1}{\sqrt{2}} & 0 & -\frac{1}{\sqrt{2}} \\
	-\frac{1}{\sqrt{6}} & \sqrt{\frac{2}{3}} & -\frac{1}{\sqrt{6}} \\
	\frac{1}{\sqrt{3}} & \frac{1}{\sqrt{3}} & \frac{1}{\sqrt{3}}
\end{pmatrix}
\begin{pmatrix}
\boldsymbol{x} \\ \boldsymbol{y} \\ \boldsymbol{z}\end{pmatrix}
\end{equation}

\begin{figure}
\centering
\includegraphics[width=0.45\linewidth]{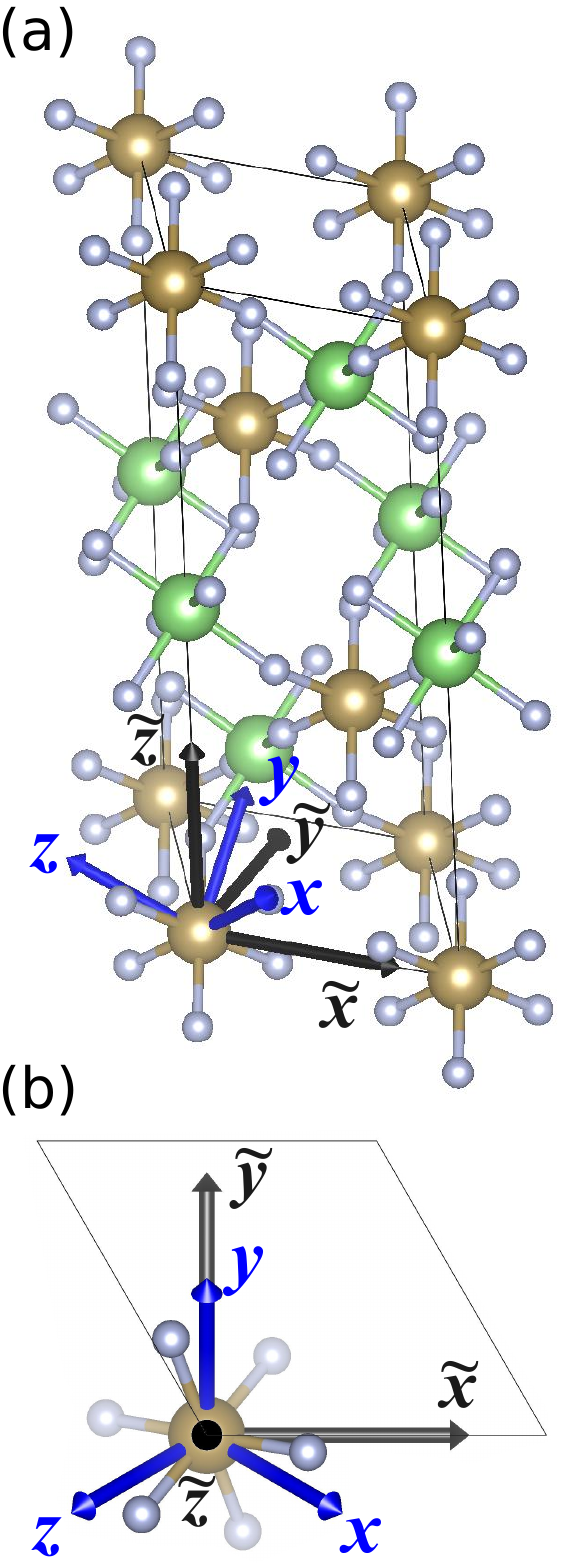}
\label{fig:coordinates}
\caption{The coordinate axes used for the double perovskites. The cartesian coordinate axes that coincide with the simple cubic (or the conventional body-centered cubic) cell are shown in blue with labels without a tilde. The cartesian coordinate axes that align with the hexagonal unit cell of the rhombohedral $R\bar{3}$ structure are shown in black with labels with tilde. }
\end{figure}

The form of piezoresistivity in the cubic phase expressed in the cartesian axes of the hexagonal system ($\tilde\Pi^{m\bar{3}m}$) can be expressed in terms of the components of $\Pi$ in the cartesian axes of the cubic phase (equation 5 in the main text) as
\begin{equation}
    \tilde{\Pi}^{m\bar{3}m} = 
\end{equation}
\makebox[\textwidth][c]{
\resizebox{1.25\textwidth}{!}{
$
    \begin{pmatrix}
         \frac{1}{2} \left(\Pi _{11}+\Pi _{12}+2 \Pi _{44}\right) & \frac{1}{6} \left(\Pi _{11}+5 \Pi _{12}-2 \Pi _{44}\right) & \frac{1}{3} \left(\Pi _{11}+2 \Pi _{12}-2 \Pi _{44}\right) & \frac{-\Pi _{11}+\Pi _{12}+2 \Pi _{44}}{3 \sqrt{2}} & 0 & 0 \\
 \frac{1}{6} \left(\Pi _{11}+5 \Pi _{12}-2 \Pi _{44}\right) & \frac{1}{2} \left(\Pi _{11}+\Pi _{12}+2 \Pi _{44}\right) & \frac{1}{3} \left(\Pi _{11}+2 \Pi _{12}-2 \Pi _{44}\right) & \frac{\Pi _{11}-\Pi _{12}-2 \Pi _{44}}{3 \sqrt{2}} & 0 & 0 \\
 \frac{1}{3} \left(\Pi _{11}+2 \Pi _{12}-2 \Pi _{44}\right) & \frac{1}{3} \left(\Pi _{11}+2 \Pi _{12}-2 \Pi _{44}\right) & \frac{1}{3} \left(\Pi _{11}+2 \Pi _{12}+4 \Pi _{44}\right) & 0 & 0 & 0 \\
 \frac{-\Pi _{11}+\Pi _{12}+2 \Pi _{44}}{3 \sqrt{2}} & \frac{\Pi _{11}-\Pi _{12}-2 \Pi _{44}}{3 \sqrt{2}} & 0 & \frac{1}{3} \left(\Pi _{11}-\Pi _{12}+\Pi _{44}\right) & 0 & 0 \\
 0 & 0 & 0 & 0 & \frac{1}{3} \left(\Pi _{11}-\Pi _{12}+\Pi _{44}\right) & \frac{-\Pi _{11}+\Pi _{12}+2 \Pi _{44}}{3 \sqrt{2}} \\
 0 & 0 & 0 & 0 & \frac{-\Pi _{11}+\Pi _{12}+2 \Pi _{44}}{3 \sqrt{2}} & \frac{1}{6} \left(\Pi _{11}-\Pi _{12}+4 \Pi _{44}\right) \\
    \end{pmatrix}
$
}
}
This shows the relation between Equations (5) and (10) in the main text.

\subsection{Ferroaxial contribution to piezoresistivity in hexagonal cell's coordinate system}

We also write explicitly the change in the piezoresistivity tensor due to ferroaxial order in the hexagonal cell, which is the analogue of Eq. (9) in the main text. As the form becomes significantly more complicated we will additionally impose the relevant physical condition that $\Psi_A, \Psi_B, \Psi_C$ all have direction $(1,1,1)$. We find:
\begin{equation}
\begin{array}{l}
 \Delta\tilde{\Pi}^\mathrm{FA} _{15}=\Psi _A-2 \Psi _B+2 \Psi _C \\
 \Delta\tilde{\Pi}^\mathrm{FA} _{16}=\sqrt{2} \left(\Psi _A+\Psi _B+2 \Psi _C\right) \\
 \Delta\tilde{\Pi}^\mathrm{FA} _{25}=-\Psi _A+2 \Psi _B-2 \Psi _C \\
 \Delta\tilde{\Pi}^\mathrm{FA} _{26}=-\sqrt{2} \left(\Psi _A+\Psi _B+2 \Psi _C\right) \\
 \Delta\tilde{\Pi}^\mathrm{FA} _{45}=-\sqrt{2} \left(\Psi _A+\Psi _B-\Psi _C\right) \\
 \Delta\tilde{\Pi}^\mathrm{FA} _{46}=-2 \Psi _A+\Psi _B+2 \Psi _C \\
 \Delta\tilde{\Pi}^\mathrm{FA} _{51}=2 \Psi _A-\Psi _B-2 \Psi _C \\
 \Delta\tilde{\Pi}^\mathrm{FA} _{52}=-2 \Psi _A+\Psi _B+2 \Psi _C \\
 \Delta\tilde{\Pi}^\mathrm{FA} _{54}=\sqrt{2} \left(\Psi _A+\Psi _B-\Psi _C\right) \\
 \Delta\tilde{\Pi}^\mathrm{FA} _{61}=-\sqrt{2} \left(\Psi _A+\Psi _B+2 \Psi _C\right) \\
 \Delta\tilde{\Pi}^\mathrm{FA} _{62}=\sqrt{2} \left(\Psi _A+\Psi _B+2 \Psi _C\right) \\
 \Delta\tilde{\Pi}^\mathrm{FA} _{64}=-\Psi _A+2 \Psi _B-2 \Psi _C \\

\end{array}
\end{equation}

\section{Landau Theories}

\subsection{Coupling between axial vectors and strain}

In this section, we discuss the Landau theory of the ferroaxial order parameter ($\Gamma_4^+$) and deviatoric strain ($\Gamma_3^+$ for normal and $\Gamma_5^+$ for shear) in a cubic ($O_h$ or $m\bar{3}m$) system. Expressing the three components of the axial moment as $(\Psi_x, \Psi_y, \Psi_z)$, the two components of normal strain as $(\varepsilon_1, \varepsilon_2)=\left(\varepsilon_{zz}-\frac{1}{2}\varepsilon_{xx}-\frac{1}{2}\varepsilon_{yy}, \frac{\sqrt{3}}{2}(\varepsilon_{yy}-\varepsilon_{xx})\right)$, and the three shear strain components as $(\varepsilon_{yz}, \varepsilon_{xz}, \varepsilon_{xy})$, the Landau free energy up to fourth order in $\Psi$ and second order in strain becomes \cite{Hatch2003}: 
\begin{align}
    \mathcal{F} & = a \Psi^2 +u_1 \Psi^4 +u_2 \left(\Psi_x^2 \Psi_y^2 +\Psi_x^2 \Psi_z^2 +\Psi_y^2 \Psi_z^2 \right) + \lambda_1 \left( \Psi_x^2 (\varepsilon_1+\sqrt{3} \varepsilon_2) +\Psi_y^2 (\varepsilon_1-\sqrt{3} \varepsilon_2) -2 \Psi_z^3 \varepsilon_1 \right) \nonumber \\ 
     & + \lambda_2\left(\Psi_x \Psi_y \varepsilon_{xy} +\Psi_x \Psi_z \varepsilon_{xz} +\Psi_y \Psi_z \varepsilon_{yz} \right) + C_1 \left( \varepsilon_1^2+\varepsilon_2^2\right) + C_2 \left( \varepsilon_{xy}^2+\varepsilon_{xz}^2+\varepsilon_{yz}^2\right) 
\end{align}
where we defined $\Psi^2=\Psi_x^2+\Psi_y^2+\Psi_z^2$, and the Latin letters represent materials-specific coefficients. 
The form of this free energy expression is rather generic, and it is identical to that of polarization coupling with strain. The $\Psi$-only part (the first three terms) allows only two low-symmetry phases with either only one component of $\Psi$ nonzero, or all three components of $\Psi$ equal to each other, determined by the sign of the coefficient $u_2$. On the other hand, the coupling with strain allows breaking the rotational symmetry in different ways. For example, in the case that the energy cost of having multiple shear strain components is large due to a higher order term $~\varepsilon_{xy}^4+\varepsilon_{xz}^4+\varepsilon_{yz}^4$, then regardless of the sign of the coefficient $\lambda_2$, a phase that has two components of $\Psi$ can be stabilized. 

\subsection{Interplay of cation order, octahedral rotations, and the ferroaxial moment}

As discussed in the main text, the octahedral rotations in perovskites do not lead to macroscopic electric toroidal dipoles or ferroaxial moments, but in the B-site checkerboard cation-ordered double perovskites, the $R$-point octahedral rotation mode is folded back onto the zone center and gives rise to a ferroaxial moment. This point can be illustrated further by considering a Landau free energy expansion that takes into account the cation order as an order parameter as well. For a cubic perovskite in the space group $Pm\bar{3}m$, the axial moments transform as the 3-dimensional irrep $\Gamma_4^+$, and the out-of-phase octahedral rotations transform as the 3-dimensional $R$-point irrep $R_4^-$ if the origin is chosen to be on the A-site. We denote the order parameters of these two irreps as 3-component vectors $\boldsymbol{\Psi}$ and $\boldsymbol{\Phi}$. Using the same origin choice, the B-site $R$-point (3D checkerboard) cation order transforms as the 1-dimensional $R_2^-$ irrep, the amplitude of which we denote by $m$. (If the origin was chosen to be on the B-site instead, then the irreps for octahedral rotations and cation orders would have been $R_4^+$ and $R_1^+$, but the form of the free energy would not have changed.) The Landau free energy in terms of these three order parameters can be derived using the standard tools \cite{Hatch2003}, and it does not include any interesting second or fourth order terms. However, in third-order, it has a trilinear coupling between these three irreps: 
\begin{equation}
    \mathcal{F}\propto m \boldsymbol{\Psi}\cdot\boldsymbol{\Phi} 
\end{equation}
When there is no cation order present and translational symmetry is not broken ($m=0$), $\boldsymbol{\Psi}$ and $\boldsymbol{\Phi}$ do not couple bilinearly with each other. However, when there is cation order ($m\neq0$), this trilinear term couples $\boldsymbol{\Psi}$ and $\boldsymbol{\Phi}$ bilinearly, and a nonzero amplitude of octahedral rotations $\boldsymbol{\Phi}$ makes it necessary that $\boldsymbol{\Psi}$ becomes nonzero as well. 

Mathematically, this is the same scenario as in hybrid-improper ferroelectric Ruddlesden-Popper structures, for which two modes from the $X$ point of the body-centered tetragonal Brillouin zone couple with the $\Gamma_5^-$ polar mode at the zone center, and hence a polarization is induced when the two separate $X$ modes are nonzero.\cite{Benedek2012, Benedek2022, Li2020} It is also similar to the trilinear terms between orthogonal components of a single zone-boundary order parameter that appear often in hexagonal lattices, for example in the vanadate Kagome metals.\cite{Christensen2021, Christensen2022} A necessary but not sufficient condition for such trilinear terms to appear is that the sum of the wavevectors adds up to zero, which is satisfied in all three cases.

\section{Additional Figures}
\begin{figure}
    \centering
    \includegraphics[width=0.65\textwidth]{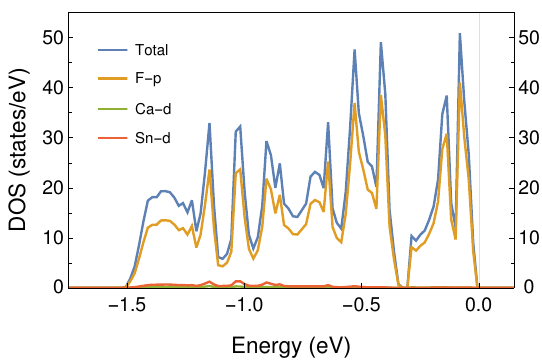}
    \caption{Density of States for the valence band showing atomic projections. The valence band has entirely Fluorine $p$ character.}
    \label{fig:DOS}
\end{figure}

Figure \ref{fig:DOS} shows the Density of States, obtained from density functional theory, of the valence band of CaSnF$_6$. Due to the large electronegativity of F, there is minimal hybridization between the cations and F, and hence the valence band is almost entirely made up of Fluorine $p$ states.

\begin{figure}
\centering
\includegraphics[width=0.65\textwidth]{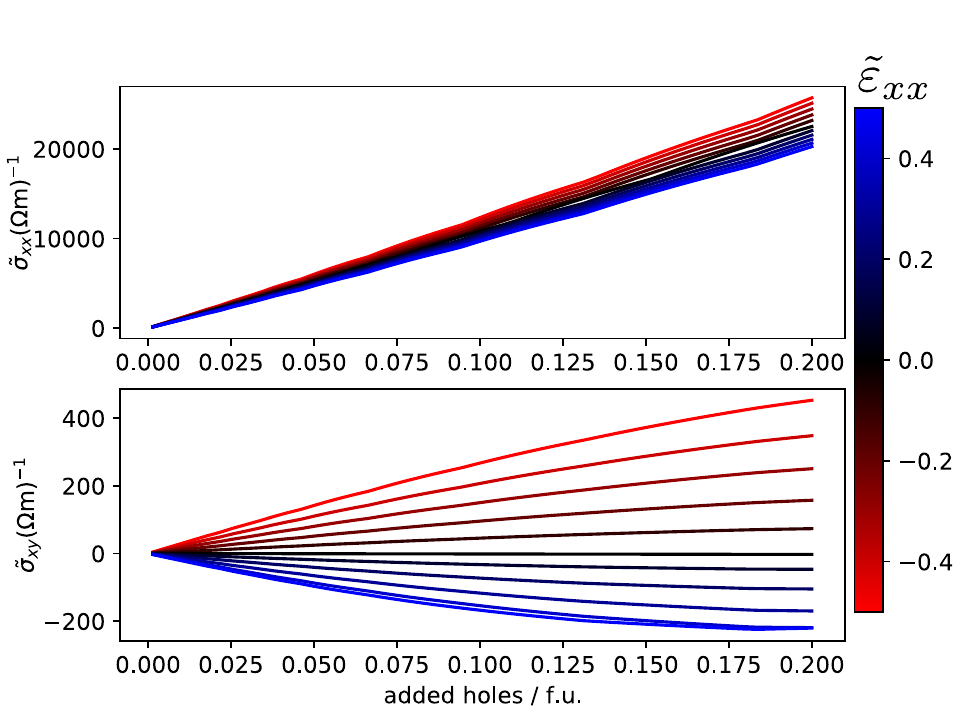}
\caption{Components of conductivity ($\sigma_{xx}$ and $\sigma_{xy}$) as a function of doping under uniaxial strain $\varepsilon_{xx}$ in the rhombohedral cell. Other components are qualitatively similar. Lines are color coded by strain. }
\label{fig:funcDoping}
\end{figure}

Fig. \ref{fig:funcDoping} shows the evolution $\sigma_{xx}$ and $\sigma_{xy}$ as a function of carrier concentration (doping), holding strain constant. The diagonal component behaves linearly and the off-diagonal non-linearly, but both are smooth with filling. 

\begin{figure}
\centering
\includegraphics[width=.65\textwidth]{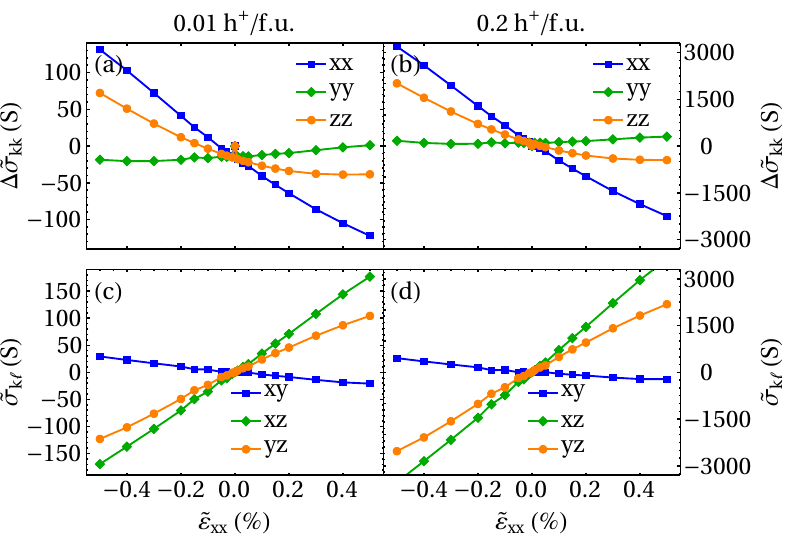}
\caption{Change in conductivity (top row) for diagonal components and absolute conductivities (bottom row) for off-diagonal components versus strain $\varepsilon_{xx}$ for two different levels of hole doping (left vs right columns) in the rhombohedral cell. The curves are very similar except for overall scale.
}
\label{fig:hexstrainvssigma}
\end{figure}

Fig. \ref{fig:hexstrainvssigma} shows the change in all the components of the conductivity under strain at two different dopings. These are nearly identical curves except at different overall scales. 

\begin{figure}
    \centering
    \includegraphics{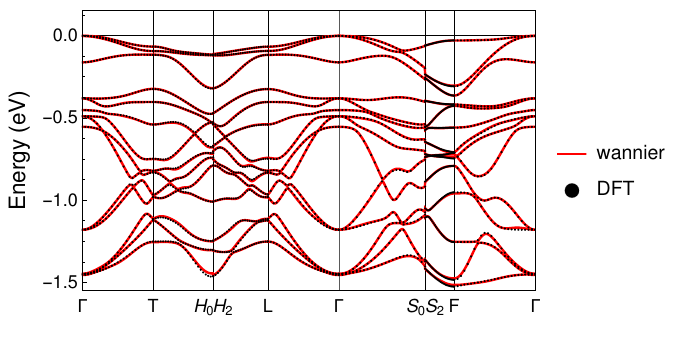}
    \caption{Band structure comparison of Wannier tight-binding model (red lines) compared and DFT energies (black dots) of valence bands. There is near perfect agreement, especially near the top of the valence band, where our calculations consider.}
    \label{fig:wannier-agreement}
\end{figure}

Fig. \ref{fig:wannier-agreement} shows that our Wannier tight-binding model accurately reproduces the entire valence band.

\begin{thebibliography}{74}%
\makeatletter
\providecommand \@ifxundefined [1]{%
 \@ifx{#1\undefined}
}%
\providecommand \@ifnum [1]{%
 \ifnum #1\expandafter \@firstoftwo
 \else \expandafter \@secondoftwo
 \fi
}%
\providecommand \@ifx [1]{%
 \ifx #1\expandafter \@firstoftwo
 \else \expandafter \@secondoftwo
 \fi
}%
\providecommand \natexlab [1]{#1}%
\providecommand \enquote  [1]{``#1''}%
\providecommand \bibnamefont  [1]{#1}%
\providecommand \bibfnamefont [1]{#1}%
\providecommand \citenamefont [1]{#1}%
\providecommand \href@noop [0]{\@secondoftwo}%
\providecommand \href [0]{\begingroup \@sanitize@url \@href}%
\providecommand \@href[1]{\@@startlink{#1}\@@href}%
\providecommand \@@href[1]{\endgroup#1\@@endlink}%
\providecommand \@sanitize@url [0]{\catcode `\\12\catcode `\$12\catcode
  `\&12\catcode `\#12\catcode `\^12\catcode `\_12\catcode `\%12\relax}%
\providecommand \@@startlink[1]{}%
\providecommand \@@endlink[0]{}%
\providecommand \url  [0]{\begingroup\@sanitize@url \@url }%
\providecommand \@url [1]{\endgroup\@href {#1}{\urlprefix }}%
\providecommand \urlprefix  [0]{URL }%
\providecommand \Eprint [0]{\href }%
\providecommand \doibase [0]{http://dx.doi.org/}%
\providecommand \selectlanguage [0]{\@gobble}%
\providecommand \bibinfo  [0]{\@secondoftwo}%
\providecommand \bibfield  [0]{\@secondoftwo}%
\providecommand \translation [1]{[#1]}%
\providecommand \BibitemOpen [0]{}%
\providecommand \bibitemStop [0]{}%
\providecommand \bibitemNoStop [0]{.\EOS\space}%
\providecommand \EOS [0]{\spacefactor3000\relax}%
\providecommand \BibitemShut  [1]{\csname bibitem#1\endcsname}%
\let\auto@bib@innerbib\@empty
\bibitem [{\citenamefont {Hayami}\ \emph {et~al.}(2018)\citenamefont {Hayami},
  \citenamefont {Yatsushiro}, \citenamefont {Yanagi},\ and\ \citenamefont
  {Kusunose}}]{Hayami2018a}%
  \BibitemOpen
  \bibfield  {author} {\bibinfo {author} {\bibfnamefont {S.}~\bibnamefont
  {Hayami}}, \bibinfo {author} {\bibfnamefont {M.}~\bibnamefont {Yatsushiro}},
  \bibinfo {author} {\bibfnamefont {Y.}~\bibnamefont {Yanagi}}, \ and\ \bibinfo
  {author} {\bibfnamefont {H.}~\bibnamefont {Kusunose}},\ }\href {\doibase
  10.1103/PhysRevB.98.165110} {\bibfield  {journal} {\bibinfo  {journal}
  {Physical Review B}\ }\textbf {\bibinfo {volume} {98}},\ \bibinfo {pages}
  {165110} (\bibinfo {year} {2018})}\BibitemShut {NoStop}%
\bibitem [{\citenamefont {Hayami}\ and\ \citenamefont
  {Kusunose}(2018)}]{Hayami2018b}%
  \BibitemOpen
  \bibfield  {author} {\bibinfo {author} {\bibfnamefont {S.}~\bibnamefont
  {Hayami}}\ and\ \bibinfo {author} {\bibfnamefont {H.}~\bibnamefont
  {Kusunose}},\ }\href {\doibase 10.7566/JPSJ.87.033709} {\bibfield  {journal}
  {\bibinfo  {journal} {Journal of the Physical Society of Japan}\ }\textbf
  {\bibinfo {volume} {87}},\ \bibinfo {pages} {033709} (\bibinfo {year}
  {2018})},\ \Eprint {http://arxiv.org/abs/1712.02927} {1712.02927}
  \BibitemShut {NoStop}%
\bibitem [{\citenamefont {Litvin}(2008)}]{Litvin2008}%
  \BibitemOpen
  \bibfield  {author} {\bibinfo {author} {\bibfnamefont {D.~B.}\ \bibnamefont
  {Litvin}},\ }\href {\doibase 10.1107/S0108767307068262} {\bibfield  {journal}
  {\bibinfo  {journal} {Acta Crystallographica Section A: Foundations of
  Crystallography}\ }\textbf {\bibinfo {volume} {64}},\ \bibinfo {pages} {316}
  (\bibinfo {year} {2008})}\BibitemShut {NoStop}%
\bibitem [{\citenamefont {Hlinka}(2014)}]{Hlinka2014}%
  \BibitemOpen
  \bibfield  {author} {\bibinfo {author} {\bibfnamefont {J.}~\bibnamefont
  {Hlinka}},\ }\href {\doibase 10.1103/PhysRevLett.113.165502} {\bibfield
  {journal} {\bibinfo  {journal} {Phys. Rev. Lett.}\ }\textbf {\bibinfo
  {volume} {113}},\ \bibinfo {pages} {165502} (\bibinfo {year}
  {2014})}\BibitemShut {NoStop}%
\bibitem [{\citenamefont {Hlinka}\ \emph {et~al.}(2016)\citenamefont {Hlinka},
  \citenamefont {Privratska}, \citenamefont {Ondrejkovic},\ and\ \citenamefont
  {Janovec}}]{Hlinka2016}%
  \BibitemOpen
  \bibfield  {author} {\bibinfo {author} {\bibfnamefont {J.}~\bibnamefont
  {Hlinka}}, \bibinfo {author} {\bibfnamefont {J.}~\bibnamefont {Privratska}},
  \bibinfo {author} {\bibfnamefont {P.}~\bibnamefont {Ondrejkovic}}, \ and\
  \bibinfo {author} {\bibfnamefont {V.}~\bibnamefont {Janovec}},\ }\href
  {\doibase 10.1103/PhysRevLett.116.177602} {\bibfield  {journal} {\bibinfo
  {journal} {Phys. Rev. Lett.}\ }\textbf {\bibinfo {volume} {116}},\ \bibinfo
  {pages} {177602} (\bibinfo {year} {2016})}\BibitemShut {NoStop}%
\bibitem [{\citenamefont {Kusunose}\ and\ \citenamefont
  {Hayami}(2024)}]{Kusunose2024}%
  \BibitemOpen
  \bibfield  {author} {\bibinfo {author} {\bibfnamefont {H.}~\bibnamefont
  {Kusunose}}\ and\ \bibinfo {author} {\bibfnamefont {S.}~\bibnamefont
  {Hayami}},\ }\href {http://arxiv.org/abs/2403.09492} {\  (\bibinfo {year}
  {2024})},\ \Eprint {http://arxiv.org/abs/2403.09492} {arXiv:2403.09492}
  \BibitemShut {NoStop}%
\bibitem [{\citenamefont {Ikeda}\ \emph {et~al.}(2015)\citenamefont {Ikeda},
  \citenamefont {Nagata}, \citenamefont {Kano},\ and\ \citenamefont
  {Mori}}]{Ikeda2015}%
  \BibitemOpen
  \bibfield  {author} {\bibinfo {author} {\bibfnamefont {N.}~\bibnamefont
  {Ikeda}}, \bibinfo {author} {\bibfnamefont {T.}~\bibnamefont {Nagata}},
  \bibinfo {author} {\bibfnamefont {J.}~\bibnamefont {Kano}}, \ and\ \bibinfo
  {author} {\bibfnamefont {S.}~\bibnamefont {Mori}},\ }\href {\doibase
  10.1088/0953-8984/27/5/053201} {\bibfield  {journal} {\bibinfo  {journal}
  {Journal of Physics: Condensed Matter}\ }\textbf {\bibinfo {volume} {27}},\
  \bibinfo {pages} {053201} (\bibinfo {year} {2015})}\BibitemShut {NoStop}%
\bibitem [{\citenamefont {Crama}(1981)}]{Crama1981}%
  \BibitemOpen
  \bibfield  {author} {\bibinfo {author} {\bibfnamefont {W.}~\bibnamefont
  {Crama}},\ }\href {\doibase https://doi.org/10.1016/0022-4596(81)90327-3}
  {\bibfield  {journal} {\bibinfo  {journal} {Journal of Solid State
  Chemistry}\ }\textbf {\bibinfo {volume} {39}},\ \bibinfo {pages} {168}
  (\bibinfo {year} {1981})}\BibitemShut {NoStop}%
\bibitem [{\citenamefont {Harada}(1982)}]{Harada1982}%
  \BibitemOpen
  \bibfield  {author} {\bibinfo {author} {\bibfnamefont {M.}~\bibnamefont
  {Harada}},\ }\href@noop {} {\bibfield  {journal} {\bibinfo  {journal}
  {Journal of the Physical Society of Japan}\ }\textbf {\bibinfo {volume}
  {51}},\ \bibinfo {pages} {2053} (\bibinfo {year} {1982})}\BibitemShut
  {NoStop}%
\bibitem [{\citenamefont {Harada}(1983)}]{Harada1983}%
  \BibitemOpen
  \bibfield  {author} {\bibinfo {author} {\bibfnamefont {M.}~\bibnamefont
  {Harada}},\ }\href@noop {} {\bibfield  {journal} {\bibinfo  {journal}
  {Journal of the Physical Society of Japan}\ }\textbf {\bibinfo {volume}
  {52}},\ \bibinfo {pages} {1646} (\bibinfo {year} {1983})}\BibitemShut
  {NoStop}%
\bibitem [{\citenamefont {Wang}\ \emph {et~al.}(2022)\citenamefont {Wang},
  \citenamefont {Petrides}, \citenamefont {McNamara}, \citenamefont {Hosen},
  \citenamefont {Lei}, \citenamefont {Wu}, \citenamefont {Hart}, \citenamefont
  {Lv}, \citenamefont {Yan}, \citenamefont {Xiao}, \citenamefont {Cha},
  \citenamefont {Narang}, \citenamefont {Schoop},\ and\ \citenamefont
  {Burch}}]{Burch2022}%
  \BibitemOpen
  \bibfield  {author} {\bibinfo {author} {\bibfnamefont {Y.}~\bibnamefont
  {Wang}}, \bibinfo {author} {\bibfnamefont {I.}~\bibnamefont {Petrides}},
  \bibinfo {author} {\bibfnamefont {G.}~\bibnamefont {McNamara}}, \bibinfo
  {author} {\bibfnamefont {M.~M.}\ \bibnamefont {Hosen}}, \bibinfo {author}
  {\bibfnamefont {S.}~\bibnamefont {Lei}}, \bibinfo {author} {\bibfnamefont
  {Y.-C.}\ \bibnamefont {Wu}}, \bibinfo {author} {\bibfnamefont {J.~L.}\
  \bibnamefont {Hart}}, \bibinfo {author} {\bibfnamefont {H.}~\bibnamefont
  {Lv}}, \bibinfo {author} {\bibfnamefont {J.}~\bibnamefont {Yan}}, \bibinfo
  {author} {\bibfnamefont {D.}~\bibnamefont {Xiao}}, \bibinfo {author}
  {\bibfnamefont {J.~J.}\ \bibnamefont {Cha}}, \bibinfo {author} {\bibfnamefont
  {P.}~\bibnamefont {Narang}}, \bibinfo {author} {\bibfnamefont {L.~M.}\
  \bibnamefont {Schoop}}, \ and\ \bibinfo {author} {\bibfnamefont {K.~S.}\
  \bibnamefont {Burch}},\ }\href {\doibase 10.1038/s41586-022-04746-6}
  {\bibfield  {journal} {\bibinfo  {journal} {Nature}\ }\textbf {\bibinfo
  {volume} {606}},\ \bibinfo {pages} {896} (\bibinfo {year}
  {2022})}\BibitemShut {NoStop}%
\bibitem [{\citenamefont {Perks}\ \emph {et~al.}(2012)\citenamefont {Perks},
  \citenamefont {Johnson}, \citenamefont {Martin}, \citenamefont {Chapon},\
  and\ \citenamefont {Radaelli}}]{Perks2012}%
  \BibitemOpen
  \bibfield  {author} {\bibinfo {author} {\bibfnamefont {N.}~\bibnamefont
  {Perks}}, \bibinfo {author} {\bibfnamefont {R.}~\bibnamefont {Johnson}},
  \bibinfo {author} {\bibfnamefont {C.}~\bibnamefont {Martin}}, \bibinfo
  {author} {\bibfnamefont {L.}~\bibnamefont {Chapon}}, \ and\ \bibinfo {author}
  {\bibfnamefont {P.}~\bibnamefont {Radaelli}},\ }\href {\doibase
  10.1038/ncomms2294} {\bibfield  {journal} {\bibinfo  {journal} {Nature
  Communications}\ }\textbf {\bibinfo {volume} {3}},\ \bibinfo {pages} {1277}
  (\bibinfo {year} {2012})}\BibitemShut {NoStop}%
\bibitem [{\citenamefont {Johnson}\ \emph {et~al.}(2012)\citenamefont
  {Johnson}, \citenamefont {Chapon}, \citenamefont {Khalyavin}, \citenamefont
  {Manuel}, \citenamefont {Radaelli},\ and\ \citenamefont
  {Martin}}]{Johnson2012}%
  \BibitemOpen
  \bibfield  {author} {\bibinfo {author} {\bibfnamefont {R.~D.}\ \bibnamefont
  {Johnson}}, \bibinfo {author} {\bibfnamefont {L.~C.}\ \bibnamefont {Chapon}},
  \bibinfo {author} {\bibfnamefont {D.~D.}\ \bibnamefont {Khalyavin}}, \bibinfo
  {author} {\bibfnamefont {P.}~\bibnamefont {Manuel}}, \bibinfo {author}
  {\bibfnamefont {P.~G.}\ \bibnamefont {Radaelli}}, \ and\ \bibinfo {author}
  {\bibfnamefont {C.}~\bibnamefont {Martin}},\ }\href {\doibase
  10.1103/PhysRevLett.108.067201} {\bibfield  {journal} {\bibinfo  {journal}
  {Physical Review Letters}\ }\textbf {\bibinfo {volume} {108}},\ \bibinfo
  {pages} {067201} (\bibinfo {year} {2012})}\BibitemShut {NoStop}%
\bibitem [{\citenamefont {Naliniand}\ and\ \citenamefont
  {Guru~Row}(2002)}]{naliniand2002}%
  \BibitemOpen
  \bibfield  {author} {\bibinfo {author} {\bibfnamefont {G.}~\bibnamefont
  {Naliniand}}\ and\ \bibinfo {author} {\bibfnamefont {T.}~\bibnamefont
  {Guru~Row}},\ }\href@noop {} {\bibfield  {journal} {\bibinfo  {journal}
  {Chemistry of materials}\ }\textbf {\bibinfo {volume} {14}},\ \bibinfo
  {pages} {4729} (\bibinfo {year} {2002})}\BibitemShut {NoStop}%
\bibitem [{\citenamefont {Wa{\'{s}}kowska}\ \emph {et~al.}(2010)\citenamefont
  {Wa{\'{s}}kowska}, \citenamefont {Gerward}, \citenamefont {{Staun Olsen}},
  \citenamefont {Morgenroth}, \citenamefont {M{\c{a}}czka},\ and\ \citenamefont
  {Hermanowicz}}]{Waskowska2010}%
  \BibitemOpen
  \bibfield  {author} {\bibinfo {author} {\bibfnamefont {A.}~\bibnamefont
  {Wa{\'{s}}kowska}}, \bibinfo {author} {\bibfnamefont {L.}~\bibnamefont
  {Gerward}}, \bibinfo {author} {\bibfnamefont {J.}~\bibnamefont {{Staun
  Olsen}}}, \bibinfo {author} {\bibfnamefont {W.}~\bibnamefont {Morgenroth}},
  \bibinfo {author} {\bibfnamefont {M.}~\bibnamefont {M{\c{a}}czka}}, \ and\
  \bibinfo {author} {\bibfnamefont {K.}~\bibnamefont {Hermanowicz}},\ }\href
  {\doibase 10.1088/0953-8984/22/5/055406} {\bibfield  {journal} {\bibinfo
  {journal} {Journal of Physics: Condensed Matter}\ }\textbf {\bibinfo {volume}
  {22}},\ \bibinfo {pages} {055406} (\bibinfo {year} {2010})}\BibitemShut
  {NoStop}%
\bibitem [{\citenamefont {Hayashida}\ \emph {et~al.}(2020)\citenamefont
  {Hayashida}, \citenamefont {Uemura}, \citenamefont {Kimura}, \citenamefont
  {Matsuoka}, \citenamefont {Morikawa}, \citenamefont {Hirose}, \citenamefont
  {Tsuda}, \citenamefont {Hasegawa},\ and\ \citenamefont
  {Kimura}}]{Hayashida2020}%
  \BibitemOpen
  \bibfield  {author} {\bibinfo {author} {\bibfnamefont {T.}~\bibnamefont
  {Hayashida}}, \bibinfo {author} {\bibfnamefont {Y.}~\bibnamefont {Uemura}},
  \bibinfo {author} {\bibfnamefont {K.}~\bibnamefont {Kimura}}, \bibinfo
  {author} {\bibfnamefont {S.}~\bibnamefont {Matsuoka}}, \bibinfo {author}
  {\bibfnamefont {D.}~\bibnamefont {Morikawa}}, \bibinfo {author}
  {\bibfnamefont {S.}~\bibnamefont {Hirose}}, \bibinfo {author} {\bibfnamefont
  {K.}~\bibnamefont {Tsuda}}, \bibinfo {author} {\bibfnamefont
  {T.}~\bibnamefont {Hasegawa}}, \ and\ \bibinfo {author} {\bibfnamefont
  {T.}~\bibnamefont {Kimura}},\ }\href {\doibase 10.1038/s41467-020-18408-6}
  {\bibfield  {journal} {\bibinfo  {journal} {Nature Communications}\ }\textbf
  {\bibinfo {volume} {11}},\ \bibinfo {pages} {4582} (\bibinfo {year}
  {2020})}\BibitemShut {NoStop}%
\bibitem [{\citenamefont {Hayashida}\ \emph {et~al.}(2021)\citenamefont
  {Hayashida}, \citenamefont {Uemura}, \citenamefont {Kimura}, \citenamefont
  {Matsuoka}, \citenamefont {Hagihala}, \citenamefont {Hirose}, \citenamefont
  {Morioka}, \citenamefont {Hasegawa},\ and\ \citenamefont
  {Kimura}}]{Hayashida2021}%
  \BibitemOpen
  \bibfield  {author} {\bibinfo {author} {\bibfnamefont {T.}~\bibnamefont
  {Hayashida}}, \bibinfo {author} {\bibfnamefont {Y.}~\bibnamefont {Uemura}},
  \bibinfo {author} {\bibfnamefont {K.}~\bibnamefont {Kimura}}, \bibinfo
  {author} {\bibfnamefont {S.}~\bibnamefont {Matsuoka}}, \bibinfo {author}
  {\bibfnamefont {M.}~\bibnamefont {Hagihala}}, \bibinfo {author}
  {\bibfnamefont {S.}~\bibnamefont {Hirose}}, \bibinfo {author} {\bibfnamefont
  {H.}~\bibnamefont {Morioka}}, \bibinfo {author} {\bibfnamefont
  {T.}~\bibnamefont {Hasegawa}}, \ and\ \bibinfo {author} {\bibfnamefont
  {T.}~\bibnamefont {Kimura}},\ }\href {\doibase
  10.1103/PhysRevMaterials.5.124409} {\bibfield  {journal} {\bibinfo  {journal}
  {Physical Review Materials}\ }\textbf {\bibinfo {volume} {5}},\ \bibinfo
  {pages} {124409} (\bibinfo {year} {2021})}\BibitemShut {NoStop}%
\bibitem [{\citenamefont {Kung}\ \emph {et~al.}(2015)\citenamefont {Kung},
  \citenamefont {Baumbach}, \citenamefont {Bauer}, \citenamefont
  {Thorsm{\o}lle}, \citenamefont {Zhang}, \citenamefont {Haule}, \citenamefont
  {Mydosh},\ and\ \citenamefont {Blumberg}}]{Kung2015}%
  \BibitemOpen
  \bibfield  {author} {\bibinfo {author} {\bibfnamefont {H.-H.}\ \bibnamefont
  {Kung}}, \bibinfo {author} {\bibfnamefont {R.~E.}\ \bibnamefont {Baumbach}},
  \bibinfo {author} {\bibfnamefont {E.~D.}\ \bibnamefont {Bauer}}, \bibinfo
  {author} {\bibfnamefont {V.~K.}\ \bibnamefont {Thorsm{\o}lle}}, \bibinfo
  {author} {\bibfnamefont {W.-L.}\ \bibnamefont {Zhang}}, \bibinfo {author}
  {\bibfnamefont {K.}~\bibnamefont {Haule}}, \bibinfo {author} {\bibfnamefont
  {J.~A.}\ \bibnamefont {Mydosh}}, \ and\ \bibinfo {author} {\bibfnamefont
  {G.}~\bibnamefont {Blumberg}},\ }\href {\doibase 10.1126/science.1259729}
  {\bibfield  {journal} {\bibinfo  {journal} {Science}\ }\textbf {\bibinfo
  {volume} {347}},\ \bibinfo {pages} {1339} (\bibinfo {year}
  {2015})}\BibitemShut {NoStop}%
\bibitem [{\citenamefont {Inda}\ \emph {et~al.}(2024)\citenamefont {Inda},
  \citenamefont {Oiwa}, \citenamefont {Hayami}, \citenamefont {Yamamoto},\ and\
  \citenamefont {Kusunose}}]{Inda2024}%
  \BibitemOpen
  \bibfield  {author} {\bibinfo {author} {\bibfnamefont {A.}~\bibnamefont
  {Inda}}, \bibinfo {author} {\bibfnamefont {R.}~\bibnamefont {Oiwa}}, \bibinfo
  {author} {\bibfnamefont {S.}~\bibnamefont {Hayami}}, \bibinfo {author}
  {\bibfnamefont {H.~M.}\ \bibnamefont {Yamamoto}}, \ and\ \bibinfo {author}
  {\bibfnamefont {H.}~\bibnamefont {Kusunose}},\ }\href@noop {} {\enquote
  {\bibinfo {title} {Quantification of chirality based on electric toroidal
  monopole},}\ } (\bibinfo {year} {2024}),\ \Eprint
  {http://arxiv.org/abs/2402.13611} {arXiv:2402.13611} \BibitemShut {NoStop}%
\bibitem [{\citenamefont {Kirikoshi}\ and\ \citenamefont
  {Hayami}(2023)}]{Kirikoshi2023}%
  \BibitemOpen
  \bibfield  {author} {\bibinfo {author} {\bibfnamefont {A.}~\bibnamefont
  {Kirikoshi}}\ and\ \bibinfo {author} {\bibfnamefont {S.}~\bibnamefont
  {Hayami}},\ }\href {\doibase 10.7566/JPSJ.92.123703} {\bibfield  {journal}
  {\bibinfo  {journal} {Journal of the Physical Society of Japan}\ }\textbf
  {\bibinfo {volume} {92}},\ \bibinfo {pages} {123703} (\bibinfo {year}
  {2023})}\BibitemShut {NoStop}%
\bibitem [{\citenamefont {Hayami}\ \emph {et~al.}(2023)\citenamefont {Hayami},
  \citenamefont {Oiwa},\ and\ \citenamefont {Kusunose}}]{Hayami2023Unconv}%
  \BibitemOpen
  \bibfield  {author} {\bibinfo {author} {\bibfnamefont {S.}~\bibnamefont
  {Hayami}}, \bibinfo {author} {\bibfnamefont {R.}~\bibnamefont {Oiwa}}, \ and\
  \bibinfo {author} {\bibfnamefont {H.}~\bibnamefont {Kusunose}},\ }\href
  {\doibase 10.1103/PhysRevB.108.085124} {\bibfield  {journal} {\bibinfo
  {journal} {Phys. Rev. B}\ }\textbf {\bibinfo {volume} {108}},\ \bibinfo
  {pages} {085124} (\bibinfo {year} {2023})}\BibitemShut {NoStop}%
\bibitem [{\citenamefont {Roy}\ \emph {et~al.}(2022)\citenamefont {Roy},
  \citenamefont {Guimar\~aes},\ and\ \citenamefont
  {S\l{}awi\ifmmode~\acute{n}\else \'{n}\fi{}ska}}]{Roy2022}%
  \BibitemOpen
  \bibfield  {author} {\bibinfo {author} {\bibfnamefont {A.}~\bibnamefont
  {Roy}}, \bibinfo {author} {\bibfnamefont {M.~H.~D.}\ \bibnamefont
  {Guimar\~aes}}, \ and\ \bibinfo {author} {\bibfnamefont {J.}~\bibnamefont
  {S\l{}awi\ifmmode~\acute{n}\else \'{n}\fi{}ska}},\ }\href {\doibase
  10.1103/PhysRevMaterials.6.045004} {\bibfield  {journal} {\bibinfo  {journal}
  {Phys. Rev. Mater.}\ }\textbf {\bibinfo {volume} {6}},\ \bibinfo {pages}
  {045004} (\bibinfo {year} {2022})}\BibitemShut {NoStop}%
\bibitem [{\citenamefont {Guo}\ \emph {et~al.}(2023)\citenamefont {Guo},
  \citenamefont {Owen}, \citenamefont {Kaczmarek}, \citenamefont {Fang},
  \citenamefont {De}, \citenamefont {Ahn}, \citenamefont {Hu}, \citenamefont
  {Agarwal}, \citenamefont {Sung}, \citenamefont {Hovden}, \citenamefont
  {Cheong},\ and\ \citenamefont {Zhao}}]{Guo2023}%
  \BibitemOpen
  \bibfield  {author} {\bibinfo {author} {\bibfnamefont {X.}~\bibnamefont
  {Guo}}, \bibinfo {author} {\bibfnamefont {R.}~\bibnamefont {Owen}}, \bibinfo
  {author} {\bibfnamefont {A.}~\bibnamefont {Kaczmarek}}, \bibinfo {author}
  {\bibfnamefont {X.}~\bibnamefont {Fang}}, \bibinfo {author} {\bibfnamefont
  {C.}~\bibnamefont {De}}, \bibinfo {author} {\bibfnamefont {Y.}~\bibnamefont
  {Ahn}}, \bibinfo {author} {\bibfnamefont {W.}~\bibnamefont {Hu}}, \bibinfo
  {author} {\bibfnamefont {N.}~\bibnamefont {Agarwal}}, \bibinfo {author}
  {\bibfnamefont {S.~H.}\ \bibnamefont {Sung}}, \bibinfo {author}
  {\bibfnamefont {R.}~\bibnamefont {Hovden}}, \bibinfo {author} {\bibfnamefont
  {S.-W.}\ \bibnamefont {Cheong}}, \ and\ \bibinfo {author} {\bibfnamefont
  {L.}~\bibnamefont {Zhao}},\ }\href {\doibase 10.1103/PhysRevB.107.L180102}
  {\bibfield  {journal} {\bibinfo  {journal} {Physical Review B}\ }\textbf
  {\bibinfo {volume} {107}},\ \bibinfo {pages} {L180102} (\bibinfo {year}
  {2023})}\BibitemShut {NoStop}%
\bibitem [{\citenamefont {Jin}\ \emph {et~al.}(2020)\citenamefont {Jin},
  \citenamefont {Drueke}, \citenamefont {Li}, \citenamefont {Admasu},
  \citenamefont {Owen}, \citenamefont {Day}, \citenamefont {Sun}, \citenamefont
  {Cheong},\ and\ \citenamefont {Zhao}}]{Jin2020}%
  \BibitemOpen
  \bibfield  {author} {\bibinfo {author} {\bibfnamefont {W.}~\bibnamefont
  {Jin}}, \bibinfo {author} {\bibfnamefont {E.}~\bibnamefont {Drueke}},
  \bibinfo {author} {\bibfnamefont {S.}~\bibnamefont {Li}}, \bibinfo {author}
  {\bibfnamefont {A.}~\bibnamefont {Admasu}}, \bibinfo {author} {\bibfnamefont
  {R.}~\bibnamefont {Owen}}, \bibinfo {author} {\bibfnamefont {M.}~\bibnamefont
  {Day}}, \bibinfo {author} {\bibfnamefont {K.}~\bibnamefont {Sun}}, \bibinfo
  {author} {\bibfnamefont {S.~W.}\ \bibnamefont {Cheong}}, \ and\ \bibinfo
  {author} {\bibfnamefont {L.}~\bibnamefont {Zhao}},\ }\href {\doibase
  10.1038/s41567-019-0695-1} {\bibfield  {journal} {\bibinfo  {journal} {Nature
  Physics}\ }\textbf {\bibinfo {volume} {16}},\ \bibinfo {pages} {42} (\bibinfo
  {year} {2020})}\BibitemShut {NoStop}%
\bibitem [{\citenamefont {Chen}\ \emph {et~al.}(2020)\citenamefont {Chen},
  \citenamefont {Wang}, \citenamefont {Xiao}, \citenamefont {Guo},
  \citenamefont {Niu},\ and\ \citenamefont {MacDonald}}]{Chen2020hall}%
  \BibitemOpen
  \bibfield  {author} {\bibinfo {author} {\bibfnamefont {H.}~\bibnamefont
  {Chen}}, \bibinfo {author} {\bibfnamefont {T.-C.}\ \bibnamefont {Wang}},
  \bibinfo {author} {\bibfnamefont {D.}~\bibnamefont {Xiao}}, \bibinfo {author}
  {\bibfnamefont {G.-Y.}\ \bibnamefont {Guo}}, \bibinfo {author} {\bibfnamefont
  {Q.}~\bibnamefont {Niu}}, \ and\ \bibinfo {author} {\bibfnamefont {A.~H.}\
  \bibnamefont {MacDonald}},\ }\href {\doibase 10.1103/PhysRevB.101.104418}
  {\bibfield  {journal} {\bibinfo  {journal} {Phys. Rev. B}\ }\textbf {\bibinfo
  {volume} {101}},\ \bibinfo {pages} {104418} (\bibinfo {year}
  {2020})}\BibitemShut {NoStop}%
\bibitem [{\citenamefont {Shao}\ \emph {et~al.}(2020)\citenamefont {Shao},
  \citenamefont {Zhang}, \citenamefont {Gurung}, \citenamefont {Yang},\ and\
  \citenamefont {Tsymbal}}]{Shao2020}%
  \BibitemOpen
  \bibfield  {author} {\bibinfo {author} {\bibfnamefont {D.-F.}\ \bibnamefont
  {Shao}}, \bibinfo {author} {\bibfnamefont {S.-H.}\ \bibnamefont {Zhang}},
  \bibinfo {author} {\bibfnamefont {G.}~\bibnamefont {Gurung}}, \bibinfo
  {author} {\bibfnamefont {W.}~\bibnamefont {Yang}}, \ and\ \bibinfo {author}
  {\bibfnamefont {E.~Y.}\ \bibnamefont {Tsymbal}},\ }\href {\doibase
  10.1103/PhysRevLett.124.067203} {\bibfield  {journal} {\bibinfo  {journal}
  {Phys. Rev. Lett.}\ }\textbf {\bibinfo {volume} {124}},\ \bibinfo {pages}
  {067203} (\bibinfo {year} {2020})}\BibitemShut {NoStop}%
\bibitem [{\citenamefont {Chu}\ \emph {et~al.}(2010)\citenamefont {Chu},
  \citenamefont {Analytis}, \citenamefont {De~Greve}, \citenamefont {McMahon},
  \citenamefont {Islam}, \citenamefont {Yamamoto},\ and\ \citenamefont
  {Fisher}}]{Chu2010}%
  \BibitemOpen
  \bibfield  {author} {\bibinfo {author} {\bibfnamefont {J.-H.}\ \bibnamefont
  {Chu}}, \bibinfo {author} {\bibfnamefont {J.~G.}\ \bibnamefont {Analytis}},
  \bibinfo {author} {\bibfnamefont {K.}~\bibnamefont {De~Greve}}, \bibinfo
  {author} {\bibfnamefont {P.~L.}\ \bibnamefont {McMahon}}, \bibinfo {author}
  {\bibfnamefont {Z.}~\bibnamefont {Islam}}, \bibinfo {author} {\bibfnamefont
  {Y.}~\bibnamefont {Yamamoto}}, \ and\ \bibinfo {author} {\bibfnamefont
  {I.~R.}\ \bibnamefont {Fisher}},\ }\href@noop {} {\bibfield  {journal}
  {\bibinfo  {journal} {Science}\ }\textbf {\bibinfo {volume} {329}},\ \bibinfo
  {pages} {824} (\bibinfo {year} {2010})}\BibitemShut {NoStop}%
\bibitem [{\citenamefont {Shapiro}\ \emph {et~al.}(2015)\citenamefont
  {Shapiro}, \citenamefont {Hlobil}, \citenamefont {Hristov}, \citenamefont
  {Maharaj},\ and\ \citenamefont {Fisher}}]{Shapiro2015}%
  \BibitemOpen
  \bibfield  {author} {\bibinfo {author} {\bibfnamefont {M.~C.}\ \bibnamefont
  {Shapiro}}, \bibinfo {author} {\bibfnamefont {P.}~\bibnamefont {Hlobil}},
  \bibinfo {author} {\bibfnamefont {A.~T.}\ \bibnamefont {Hristov}}, \bibinfo
  {author} {\bibfnamefont {A.~V.}\ \bibnamefont {Maharaj}}, \ and\ \bibinfo
  {author} {\bibfnamefont {I.~R.}\ \bibnamefont {Fisher}},\ }\href {\doibase
  10.1103/PhysRevB.92.235147} {\bibfield  {journal} {\bibinfo  {journal} {Phys.
  Rev. B}\ }\textbf {\bibinfo {volume} {92}},\ \bibinfo {pages} {235147}
  (\bibinfo {year} {2015})}\BibitemShut {NoStop}%
\bibitem [{\citenamefont {Fernandes}\ and\ \citenamefont
  {Schmalian}(2012)}]{Fernandes2012}%
  \BibitemOpen
  \bibfield  {author} {\bibinfo {author} {\bibfnamefont {R.~M.}\ \bibnamefont
  {Fernandes}}\ and\ \bibinfo {author} {\bibfnamefont {J.}~\bibnamefont
  {Schmalian}},\ }\href {\doibase 10.1088/0953-2048/25/8/084005} {\bibfield
  {journal} {\bibinfo  {journal} {Superconductor Science and Technology}\
  }\textbf {\bibinfo {volume} {25}},\ \bibinfo {pages} {084005} (\bibinfo
  {year} {2012})}\BibitemShut {NoStop}%
\bibitem [{\citenamefont {Fradkin}\ \emph {et~al.}(2010)\citenamefont
  {Fradkin}, \citenamefont {Kivelson}, \citenamefont {Lawler}, \citenamefont
  {Eisenstein},\ and\ \citenamefont {Mackenzie}}]{Fradkin2010}%
  \BibitemOpen
  \bibfield  {author} {\bibinfo {author} {\bibfnamefont {E.}~\bibnamefont
  {Fradkin}}, \bibinfo {author} {\bibfnamefont {S.~A.}\ \bibnamefont
  {Kivelson}}, \bibinfo {author} {\bibfnamefont {M.~J.}\ \bibnamefont
  {Lawler}}, \bibinfo {author} {\bibfnamefont {J.~P.}\ \bibnamefont
  {Eisenstein}}, \ and\ \bibinfo {author} {\bibfnamefont {A.~P.}\ \bibnamefont
  {Mackenzie}},\ }\href {\doibase 10.1146/annurev-conmatphys-070909-103925}
  {\bibfield  {journal} {\bibinfo  {journal} {Annual Review of Condensed Matter
  Physics}\ }\textbf {\bibinfo {volume} {1}},\ \bibinfo {pages} {153} (\bibinfo
  {year} {2010})}\BibitemShut {NoStop}%
\bibitem [{\citenamefont {Palmstrom}(2020)}]{Palmstrom2020Thesis}%
  \BibitemOpen
  \bibfield  {author} {\bibinfo {author} {\bibfnamefont {J.~C.}\ \bibnamefont
  {Palmstrom}},\ }\href@noop {} {\emph {\bibinfo {title} {Elastoresistance of
  Iron-Based Superconductors}}}\ (\bibinfo  {publisher} {Stanford University},\
  \bibinfo {year} {2020})\BibitemShut {NoStop}%
\bibitem [{\citenamefont {Chu}\ \emph {et~al.}(2012)\citenamefont {Chu},
  \citenamefont {Kuo}, \citenamefont {Analytis},\ and\ \citenamefont
  {Fisher}}]{Chu2012}%
  \BibitemOpen
  \bibfield  {author} {\bibinfo {author} {\bibfnamefont {J.-H.}\ \bibnamefont
  {Chu}}, \bibinfo {author} {\bibfnamefont {H.-H.}\ \bibnamefont {Kuo}},
  \bibinfo {author} {\bibfnamefont {J.~G.}\ \bibnamefont {Analytis}}, \ and\
  \bibinfo {author} {\bibfnamefont {I.~R.}\ \bibnamefont {Fisher}},\
  }\href@noop {} {\bibfield  {journal} {\bibinfo  {journal} {Science}\ }\textbf
  {\bibinfo {volume} {337}},\ \bibinfo {pages} {710} (\bibinfo {year}
  {2012})}\BibitemShut {NoStop}%
\bibitem [{\citenamefont {Mayer}\ \emph {et~al.}(1983)\citenamefont {Mayer},
  \citenamefont {Reinen},\ and\ \citenamefont {Heger}}]{mayer1983struktur}%
  \BibitemOpen
  \bibfield  {author} {\bibinfo {author} {\bibfnamefont {H.}~\bibnamefont
  {Mayer}}, \bibinfo {author} {\bibfnamefont {D.}~\bibnamefont {Reinen}}, \
  and\ \bibinfo {author} {\bibfnamefont {G.}~\bibnamefont {Heger}},\
  }\href@noop {} {\bibfield  {journal} {\bibinfo  {journal} {Journal of Solid
  State Chemistry}\ }\textbf {\bibinfo {volume} {50}},\ \bibinfo {pages} {213}
  (\bibinfo {year} {1983})}\BibitemShut {NoStop}%
\bibitem [{\citenamefont {Brouwer}\ and\ \citenamefont
  {Jellinek}(1980)}]{Brouwer1980}%
  \BibitemOpen
  \bibfield  {author} {\bibinfo {author} {\bibfnamefont {R.}~\bibnamefont
  {Brouwer}}\ and\ \bibinfo {author} {\bibfnamefont {F.}~\bibnamefont
  {Jellinek}},\ }\href {\doibase 10.1016/0378-4363(80)90209-0} {\bibfield
  {journal} {\bibinfo  {journal} {Physica B+C}\ }\textbf {\bibinfo {volume}
  {99}},\ \bibinfo {pages} {51} (\bibinfo {year} {1980})}\BibitemShut {NoStop}%
\bibitem [{\citenamefont {Pizzi}\ \emph
  {et~al.}(2014{\natexlab{a}})\citenamefont {Pizzi}, \citenamefont {Volja},
  \citenamefont {Kozinsky}, \citenamefont {Fornari},\ and\ \citenamefont
  {Marzari}}]{pizzi2014boltzwann}%
  \BibitemOpen
  \bibfield  {author} {\bibinfo {author} {\bibfnamefont {G.}~\bibnamefont
  {Pizzi}}, \bibinfo {author} {\bibfnamefont {D.}~\bibnamefont {Volja}},
  \bibinfo {author} {\bibfnamefont {B.}~\bibnamefont {Kozinsky}}, \bibinfo
  {author} {\bibfnamefont {M.}~\bibnamefont {Fornari}}, \ and\ \bibinfo
  {author} {\bibfnamefont {N.}~\bibnamefont {Marzari}},\ }\href@noop {}
  {\bibfield  {journal} {\bibinfo  {journal} {Computer Physics Communications}\
  }\textbf {\bibinfo {volume} {185}},\ \bibinfo {pages} {422} (\bibinfo {year}
  {2014}{\natexlab{a}})}\BibitemShut {NoStop}%
\bibitem [{\citenamefont {Riggs}\ \emph {et~al.}(2015)\citenamefont {Riggs},
  \citenamefont {Shapiro}, \citenamefont {Maharaj}, \citenamefont {Raghu},
  \citenamefont {Bauer}, \citenamefont {Baumbach}, \citenamefont
  {Giraldo-Gallo}, \citenamefont {Wartenbe},\ and\ \citenamefont
  {Fisher}}]{Riggs2015}%
  \BibitemOpen
  \bibfield  {author} {\bibinfo {author} {\bibfnamefont {S.~C.}\ \bibnamefont
  {Riggs}}, \bibinfo {author} {\bibfnamefont {M.}~\bibnamefont {Shapiro}},
  \bibinfo {author} {\bibfnamefont {A.~V.}\ \bibnamefont {Maharaj}}, \bibinfo
  {author} {\bibfnamefont {S.}~\bibnamefont {Raghu}}, \bibinfo {author}
  {\bibfnamefont {E.}~\bibnamefont {Bauer}}, \bibinfo {author} {\bibfnamefont
  {R.}~\bibnamefont {Baumbach}}, \bibinfo {author} {\bibfnamefont
  {P.}~\bibnamefont {Giraldo-Gallo}}, \bibinfo {author} {\bibfnamefont
  {M.}~\bibnamefont {Wartenbe}}, \ and\ \bibinfo {author} {\bibfnamefont
  {I.}~\bibnamefont {Fisher}},\ }\href@noop {} {\bibfield  {journal} {\bibinfo
  {journal} {Nature Communications}\ }\textbf {\bibinfo {volume} {6}},\
  \bibinfo {pages} {6425} (\bibinfo {year} {2015})}\BibitemShut {NoStop}%
\bibitem [{\citenamefont {Hosoi}\ \emph {et~al.}(2016)\citenamefont {Hosoi},
  \citenamefont {Matsuura}, \citenamefont {Ishida}, \citenamefont {Wang},
  \citenamefont {Mizukami}, \citenamefont {Watashige}, \citenamefont
  {Kasahara}, \citenamefont {Matsuda},\ and\ \citenamefont
  {Shibauchi}}]{Hosoi2016}%
  \BibitemOpen
  \bibfield  {author} {\bibinfo {author} {\bibfnamefont {S.}~\bibnamefont
  {Hosoi}}, \bibinfo {author} {\bibfnamefont {K.}~\bibnamefont {Matsuura}},
  \bibinfo {author} {\bibfnamefont {K.}~\bibnamefont {Ishida}}, \bibinfo
  {author} {\bibfnamefont {H.}~\bibnamefont {Wang}}, \bibinfo {author}
  {\bibfnamefont {Y.}~\bibnamefont {Mizukami}}, \bibinfo {author}
  {\bibfnamefont {T.}~\bibnamefont {Watashige}}, \bibinfo {author}
  {\bibfnamefont {S.}~\bibnamefont {Kasahara}}, \bibinfo {author}
  {\bibfnamefont {Y.}~\bibnamefont {Matsuda}}, \ and\ \bibinfo {author}
  {\bibfnamefont {T.}~\bibnamefont {Shibauchi}},\ }\href@noop {} {\bibfield
  {journal} {\bibinfo  {journal} {Proceedings of the National Academy of
  Sciences}\ }\textbf {\bibinfo {volume} {113}},\ \bibinfo {pages} {8139}
  (\bibinfo {year} {2016})}\BibitemShut {NoStop}%
\bibitem [{\citenamefont {Ishida}\ \emph {et~al.}(2020)\citenamefont {Ishida},
  \citenamefont {Hosoi}, \citenamefont {Teramoto}, \citenamefont {Usui},
  \citenamefont {Mizukami}, \citenamefont {Itaka}, \citenamefont {Matsuda},
  \citenamefont {Watanabe},\ and\ \citenamefont {Shibauchi}}]{Ishida2020}%
  \BibitemOpen
  \bibfield  {author} {\bibinfo {author} {\bibfnamefont {K.}~\bibnamefont
  {Ishida}}, \bibinfo {author} {\bibfnamefont {S.}~\bibnamefont {Hosoi}},
  \bibinfo {author} {\bibfnamefont {Y.}~\bibnamefont {Teramoto}}, \bibinfo
  {author} {\bibfnamefont {T.}~\bibnamefont {Usui}}, \bibinfo {author}
  {\bibfnamefont {Y.}~\bibnamefont {Mizukami}}, \bibinfo {author}
  {\bibfnamefont {K.}~\bibnamefont {Itaka}}, \bibinfo {author} {\bibfnamefont
  {Y.}~\bibnamefont {Matsuda}}, \bibinfo {author} {\bibfnamefont
  {T.}~\bibnamefont {Watanabe}}, \ and\ \bibinfo {author} {\bibfnamefont
  {T.}~\bibnamefont {Shibauchi}},\ }\href@noop {} {\bibfield  {journal}
  {\bibinfo  {journal} {Journal of the Physical Society of Japan}\ }\textbf
  {\bibinfo {volume} {89}},\ \bibinfo {pages} {064707} (\bibinfo {year}
  {2020})}\BibitemShut {NoStop}%
\bibitem [{\citenamefont {Eckberg}\ \emph {et~al.}(2020)\citenamefont
  {Eckberg}, \citenamefont {Campbell}, \citenamefont {Metz}, \citenamefont
  {Collini}, \citenamefont {Hodovanets}, \citenamefont {Drye}, \citenamefont
  {Zavalij}, \citenamefont {Christensen}, \citenamefont {Fernandes},
  \citenamefont {Lee} \emph {et~al.}}]{Eckberg2020}%
  \BibitemOpen
  \bibfield  {author} {\bibinfo {author} {\bibfnamefont {C.}~\bibnamefont
  {Eckberg}}, \bibinfo {author} {\bibfnamefont {D.~J.}\ \bibnamefont
  {Campbell}}, \bibinfo {author} {\bibfnamefont {T.}~\bibnamefont {Metz}},
  \bibinfo {author} {\bibfnamefont {J.}~\bibnamefont {Collini}}, \bibinfo
  {author} {\bibfnamefont {H.}~\bibnamefont {Hodovanets}}, \bibinfo {author}
  {\bibfnamefont {T.}~\bibnamefont {Drye}}, \bibinfo {author} {\bibfnamefont
  {P.}~\bibnamefont {Zavalij}}, \bibinfo {author} {\bibfnamefont {M.~H.}\
  \bibnamefont {Christensen}}, \bibinfo {author} {\bibfnamefont {R.~M.}\
  \bibnamefont {Fernandes}}, \bibinfo {author} {\bibfnamefont {S.}~\bibnamefont
  {Lee}},  \emph {et~al.},\ }\href@noop {} {\bibfield  {journal} {\bibinfo
  {journal} {Nature Physics}\ }\textbf {\bibinfo {volume} {16}},\ \bibinfo
  {pages} {346} (\bibinfo {year} {2020})}\BibitemShut {NoStop}%
\bibitem [{Sup()}]{Supplement}%
  \BibitemOpen
  \href@noop {} {}\bibinfo {note} {See supplemental information, which contains
  references \cite{VASP, VASP-PAW, PBEsol, Brubeval201, Aroyo2006,
  wilson1975charge, Hatch2003, Benedek2012, Benedek2022, Li2020,
  Christensen2021, Christensen2022}, for the details of first-principles
  calculations, tensor decomposition, form of the piezoresistivity tensor in
  other point groups, and the discussion of TaS$_2$.}\BibitemShut {Stop}%
\bibitem [{\citenamefont {Nye}(1985)}]{Nye1985book}%
  \BibitemOpen
  \bibfield  {author} {\bibinfo {author} {\bibfnamefont {J.~F.}\ \bibnamefont
  {Nye}},\ }\href@noop {} {\emph {\bibinfo {title} {Physical properties of
  crystals: their representation by tensors and matrices}}}\ (\bibinfo
  {publisher} {Oxford university press},\ \bibinfo {year} {1985})\BibitemShut
  {NoStop}%
\bibitem [{\citenamefont {Jahn}(1949)}]{Jahn1949}%
  \BibitemOpen
  \bibfield  {author} {\bibinfo {author} {\bibfnamefont {H.~A.}\ \bibnamefont
  {Jahn}},\ }\href {\doibase 10.1107/S0365110X49000060} {\bibfield  {journal}
  {\bibinfo  {journal} {Acta Crystallographica}\ }\textbf {\bibinfo {volume}
  {2}},\ \bibinfo {pages} {30} (\bibinfo {year} {1949})}\BibitemShut {NoStop}%
\bibitem [{\citenamefont {Dresselhaus}\ \emph {et~al.}(2008)\citenamefont
  {Dresselhaus}, \citenamefont {Dresselhaus},\ and\ \citenamefont
  {Jorio}}]{dresselhaus}%
  \BibitemOpen
  \bibfield  {author} {\bibinfo {author} {\bibfnamefont {M.~S.}\ \bibnamefont
  {Dresselhaus}}, \bibinfo {author} {\bibfnamefont {G.}~\bibnamefont
  {Dresselhaus}}, \ and\ \bibinfo {author} {\bibfnamefont {A.}~\bibnamefont
  {Jorio}},\ }\href@noop {} {\emph {\bibinfo {title} {Applications of group
  theory to the physics of solids}}}\ (\bibinfo  {publisher} {Springer
  Berlin},\ \bibinfo {year} {2008})\BibitemShut {NoStop}%
\bibitem [{Note1()}]{Note1}%
  \BibitemOpen
  \bibinfo {note} {For a system with broken time reversal symmetry, the Jahn
  symbol becomes [V$^2$]$^*$[V$^2$].}\BibitemShut {Stop}%
\bibitem [{\citenamefont {Hecker}\ \emph {et~al.}(2024)\citenamefont {Hecker},
  \citenamefont {Rastogi}, \citenamefont {Agterberg},\ and\ \citenamefont
  {Fernandes}}]{Hecker2024}%
  \BibitemOpen
  \bibfield  {author} {\bibinfo {author} {\bibfnamefont {M.}~\bibnamefont
  {Hecker}}, \bibinfo {author} {\bibfnamefont {A.}~\bibnamefont {Rastogi}},
  \bibinfo {author} {\bibfnamefont {D.~F.}\ \bibnamefont {Agterberg}}, \ and\
  \bibinfo {author} {\bibfnamefont {R.~M.}\ \bibnamefont {Fernandes}},\
  }\href@noop {} {\bibfield  {journal} {\bibinfo  {journal} {arXiv:2402.17657}\
  } (\bibinfo {year} {2024})}\BibitemShut {NoStop}%
\bibitem [{\citenamefont {Pellicer-Porres}\ \emph {et~al.}(2001)\citenamefont
  {Pellicer-Porres}, \citenamefont {Segura}, \citenamefont {Mu\~noz},
  \citenamefont {Z\'u\~niga}, \citenamefont {Iti\'e}, \citenamefont {Polian},\
  and\ \citenamefont {Munsch}}]{Pellicer-Porres2001}%
  \BibitemOpen
  \bibfield  {author} {\bibinfo {author} {\bibfnamefont {J.}~\bibnamefont
  {Pellicer-Porres}}, \bibinfo {author} {\bibfnamefont {A.}~\bibnamefont
  {Segura}}, \bibinfo {author} {\bibfnamefont {V.}~\bibnamefont {Mu\~noz}},
  \bibinfo {author} {\bibfnamefont {J.}~\bibnamefont {Z\'u\~niga}}, \bibinfo
  {author} {\bibfnamefont {J.~P.}\ \bibnamefont {Iti\'e}}, \bibinfo {author}
  {\bibfnamefont {A.}~\bibnamefont {Polian}}, \ and\ \bibinfo {author}
  {\bibfnamefont {P.}~\bibnamefont {Munsch}},\ }\href {\doibase
  10.1103/PhysRevB.65.012109} {\bibfield  {journal} {\bibinfo  {journal} {Phys.
  Rev. B}\ }\textbf {\bibinfo {volume} {65}},\ \bibinfo {pages} {012109}
  (\bibinfo {year} {2001})}\BibitemShut {NoStop}%
\bibitem [{\citenamefont {Piltz}\ \emph {et~al.}(1995)\citenamefont {Piltz},
  \citenamefont {Maclean}, \citenamefont {Clark}, \citenamefont {Ackland},
  \citenamefont {Hatton},\ and\ \citenamefont {Crain}}]{Piltz1995}%
  \BibitemOpen
  \bibfield  {author} {\bibinfo {author} {\bibfnamefont {R.~O.}\ \bibnamefont
  {Piltz}}, \bibinfo {author} {\bibfnamefont {J.~R.}\ \bibnamefont {Maclean}},
  \bibinfo {author} {\bibfnamefont {S.~J.}\ \bibnamefont {Clark}}, \bibinfo
  {author} {\bibfnamefont {G.~J.}\ \bibnamefont {Ackland}}, \bibinfo {author}
  {\bibfnamefont {P.~D.}\ \bibnamefont {Hatton}}, \ and\ \bibinfo {author}
  {\bibfnamefont {J.}~\bibnamefont {Crain}},\ }\href {\doibase
  10.1103/PhysRevB.52.4072} {\bibfield  {journal} {\bibinfo  {journal} {Phys.
  Rev. B}\ }\textbf {\bibinfo {volume} {52}},\ \bibinfo {pages} {4072}
  (\bibinfo {year} {1995})}\BibitemShut {NoStop}%
\bibitem [{\citenamefont {van Egmond}(1975)}]{vanEgmond1975}%
  \BibitemOpen
  \bibfield  {author} {\bibinfo {author} {\bibfnamefont {A.}~\bibnamefont {van
  Egmond}},\ }\href {\doibase 10.1016/0022-1902(75)80917-1} {\bibfield
  {journal} {\bibinfo  {journal} {Journal of Inorganic and Nuclear Chemistry}\
  }\textbf {\bibinfo {volume} {37}},\ \bibinfo {pages} {1929–1931} (\bibinfo
  {year} {1975})}\BibitemShut {NoStop}%
\bibitem [{\citenamefont {Liang}\ \emph {et~al.}(1989)\citenamefont {Liang},
  \citenamefont {Rao},\ and\ \citenamefont {Zhang}}]{Liang1989}%
  \BibitemOpen
  \bibfield  {author} {\bibinfo {author} {\bibfnamefont {J.~K.}\ \bibnamefont
  {Liang}}, \bibinfo {author} {\bibfnamefont {G.~H.}\ \bibnamefont {Rao}}, \
  and\ \bibinfo {author} {\bibfnamefont {Y.~M.}\ \bibnamefont {Zhang}},\ }\href
  {\doibase 10.1103/PhysRevB.39.459} {\bibfield  {journal} {\bibinfo  {journal}
  {Phys. Rev. B}\ }\textbf {\bibinfo {volume} {39}},\ \bibinfo {pages} {459}
  (\bibinfo {year} {1989})}\BibitemShut {NoStop}%
\bibitem [{\citenamefont {Jain}\ \emph {et~al.}(2013)\citenamefont {Jain},
  \citenamefont {Ong}, \citenamefont {Hautier}, \citenamefont {Chen},
  \citenamefont {Richards}, \citenamefont {Dacek}, \citenamefont {Cholia},
  \citenamefont {Gunter}, \citenamefont {Skinner}, \citenamefont {Ceder},\ and\
  \citenamefont {Persson}}]{Jain2013}%
  \BibitemOpen
  \bibfield  {author} {\bibinfo {author} {\bibfnamefont {A.}~\bibnamefont
  {Jain}}, \bibinfo {author} {\bibfnamefont {S.~P.}\ \bibnamefont {Ong}},
  \bibinfo {author} {\bibfnamefont {G.}~\bibnamefont {Hautier}}, \bibinfo
  {author} {\bibfnamefont {W.}~\bibnamefont {Chen}}, \bibinfo {author}
  {\bibfnamefont {W.~D.}\ \bibnamefont {Richards}}, \bibinfo {author}
  {\bibfnamefont {S.}~\bibnamefont {Dacek}}, \bibinfo {author} {\bibfnamefont
  {S.}~\bibnamefont {Cholia}}, \bibinfo {author} {\bibfnamefont
  {D.}~\bibnamefont {Gunter}}, \bibinfo {author} {\bibfnamefont
  {D.}~\bibnamefont {Skinner}}, \bibinfo {author} {\bibfnamefont
  {G.}~\bibnamefont {Ceder}}, \ and\ \bibinfo {author} {\bibfnamefont {K.~A.}\
  \bibnamefont {Persson}},\ }\href {\doibase 10.1063/1.4812323} {\bibfield
  {journal} {\bibinfo  {journal} {APL Materials}\ }\textbf {\bibinfo {volume}
  {1}},\ \bibinfo {pages} {011002} (\bibinfo {year} {2013})}\BibitemShut
  {NoStop}%
\bibitem [{\citenamefont {Evans}\ \emph {et~al.}(2020)\citenamefont {Evans},
  \citenamefont {Wu}, \citenamefont {Seshadri},\ and\ \citenamefont
  {Cheetham}}]{Evans2020}%
  \BibitemOpen
  \bibfield  {author} {\bibinfo {author} {\bibfnamefont {H.~A.}\ \bibnamefont
  {Evans}}, \bibinfo {author} {\bibfnamefont {Y.}~\bibnamefont {Wu}}, \bibinfo
  {author} {\bibfnamefont {R.}~\bibnamefont {Seshadri}}, \ and\ \bibinfo
  {author} {\bibfnamefont {A.~K.}\ \bibnamefont {Cheetham}},\ }\href {\doibase
  10.1038/s41578-019-0160-x} {\bibfield  {journal} {\bibinfo  {journal} {Nature
  Reviews Materials}\ }\textbf {\bibinfo {volume} {5}},\ \bibinfo {pages} {196}
  (\bibinfo {year} {2020})}\BibitemShut {NoStop}%
\bibitem [{\citenamefont {Gao}\ \emph {et~al.}(2023)\citenamefont {Gao},
  \citenamefont {Zhang}, \citenamefont {Jiao}, \citenamefont {Qiao},
  \citenamefont {Sanson}, \citenamefont {Sun}, \citenamefont {Shi},
  \citenamefont {Liang},\ and\ \citenamefont {Chen}}]{gao2023new}%
  \BibitemOpen
  \bibfield  {author} {\bibinfo {author} {\bibfnamefont {Q.}~\bibnamefont
  {Gao}}, \bibinfo {author} {\bibfnamefont {S.}~\bibnamefont {Zhang}}, \bibinfo
  {author} {\bibfnamefont {Y.}~\bibnamefont {Jiao}}, \bibinfo {author}
  {\bibfnamefont {Y.}~\bibnamefont {Qiao}}, \bibinfo {author} {\bibfnamefont
  {A.}~\bibnamefont {Sanson}}, \bibinfo {author} {\bibfnamefont
  {Q.}~\bibnamefont {Sun}}, \bibinfo {author} {\bibfnamefont {X.}~\bibnamefont
  {Shi}}, \bibinfo {author} {\bibfnamefont {E.}~\bibnamefont {Liang}}, \ and\
  \bibinfo {author} {\bibfnamefont {J.}~\bibnamefont {Chen}},\ }\href@noop {}
  {\bibfield  {journal} {\bibinfo  {journal} {Nano Research}\ }\textbf
  {\bibinfo {volume} {16}},\ \bibinfo {pages} {5964} (\bibinfo {year}
  {2023})}\BibitemShut {NoStop}%
\bibitem [{\citenamefont {Lufaso}\ and\ \citenamefont
  {Woodward}(2004)}]{Lufaso2004}%
  \BibitemOpen
  \bibfield  {author} {\bibinfo {author} {\bibfnamefont {M.~W.}\ \bibnamefont
  {Lufaso}}\ and\ \bibinfo {author} {\bibfnamefont {P.~M.}\ \bibnamefont
  {Woodward}},\ }\href {\doibase 10.1107/S0108768103026661} {\bibfield
  {journal} {\bibinfo  {journal} {Acta Crystallographica Section B: Structural
  Science}\ }\textbf {\bibinfo {volume} {60}},\ \bibinfo {pages} {10} (\bibinfo
  {year} {2004})}\BibitemShut {NoStop}%
\bibitem [{\citenamefont {Glazer}(1972)}]{Glazer1972}%
  \BibitemOpen
  \bibfield  {author} {\bibinfo {author} {\bibfnamefont {A.~M.}\ \bibnamefont
  {Glazer}},\ }\href {\doibase 10.1107/S0567740872007976} {\bibfield  {journal}
  {\bibinfo  {journal} {Acta Crystallographica Section B Structural
  Crystallography and Crystal Chemistry}\ }\textbf {\bibinfo {volume} {28}},\
  \bibinfo {pages} {3384} (\bibinfo {year} {1972})}\BibitemShut {NoStop}%
\bibitem [{\citenamefont {Goldman}(2014)}]{Goldman2014}%
  \BibitemOpen
  \bibfield  {author} {\bibinfo {author} {\bibfnamefont {A.}~\bibnamefont
  {Goldman}},\ }\href {\doibase 10.1146/annurev-matsci-070813-113407}
  {\bibfield  {journal} {\bibinfo  {journal} {Annual Review of Materials
  Research}\ }\textbf {\bibinfo {volume} {44}},\ \bibinfo {pages} {45}
  (\bibinfo {year} {2014})}\BibitemShut {NoStop}%
\bibitem [{\citenamefont {Leighton}\ \emph {et~al.}(2022)\citenamefont
  {Leighton}, \citenamefont {Birol},\ and\ \citenamefont
  {Walter}}]{Leighton2022}%
  \BibitemOpen
  \bibfield  {author} {\bibinfo {author} {\bibfnamefont {C.}~\bibnamefont
  {Leighton}}, \bibinfo {author} {\bibfnamefont {T.}~\bibnamefont {Birol}}, \
  and\ \bibinfo {author} {\bibfnamefont {J.}~\bibnamefont {Walter}},\
  }\href@noop {} {\bibfield  {journal} {\bibinfo  {journal} {APL Materials}\
  }\textbf {\bibinfo {volume} {10}},\ \bibinfo {pages} {040901} (\bibinfo
  {year} {2022})}\BibitemShut {NoStop}%
\bibitem [{\citenamefont {Pizzi}\ \emph
  {et~al.}(2014{\natexlab{b}})\citenamefont {Pizzi}, \citenamefont {Volja},
  \citenamefont {Kozinsky}, \citenamefont {Fornari},\ and\ \citenamefont
  {Marzari}}]{pizzi2014updated}%
  \BibitemOpen
  \bibfield  {author} {\bibinfo {author} {\bibfnamefont {G.}~\bibnamefont
  {Pizzi}}, \bibinfo {author} {\bibfnamefont {D.}~\bibnamefont {Volja}},
  \bibinfo {author} {\bibfnamefont {B.}~\bibnamefont {Kozinsky}}, \bibinfo
  {author} {\bibfnamefont {M.}~\bibnamefont {Fornari}}, \ and\ \bibinfo
  {author} {\bibfnamefont {N.}~\bibnamefont {Marzari}},\ }\href@noop {}
  {\bibfield  {journal} {\bibinfo  {journal} {Comput. Phys. Commun.}\ }\textbf
  {\bibinfo {volume} {185}},\ \bibinfo {pages} {2311} (\bibinfo {year}
  {2014}{\natexlab{b}})}\BibitemShut {NoStop}%
\bibitem [{\citenamefont {Mostofi}\ \emph {et~al.}(2014)\citenamefont
  {Mostofi}, \citenamefont {Yates}, \citenamefont {Pizzi}, \citenamefont {Lee},
  \citenamefont {Souza}, \citenamefont {Vanderbilt},\ and\ \citenamefont
  {Marzari}}]{wannier90}%
  \BibitemOpen
  \bibfield  {author} {\bibinfo {author} {\bibfnamefont {A.~A.}\ \bibnamefont
  {Mostofi}}, \bibinfo {author} {\bibfnamefont {J.~R.}\ \bibnamefont {Yates}},
  \bibinfo {author} {\bibfnamefont {G.}~\bibnamefont {Pizzi}}, \bibinfo
  {author} {\bibfnamefont {Y.-S.}\ \bibnamefont {Lee}}, \bibinfo {author}
  {\bibfnamefont {I.}~\bibnamefont {Souza}}, \bibinfo {author} {\bibfnamefont
  {D.}~\bibnamefont {Vanderbilt}}, \ and\ \bibinfo {author} {\bibfnamefont
  {N.}~\bibnamefont {Marzari}},\ }\href@noop {} {\bibfield  {journal} {\bibinfo
   {journal} {Computer Physics Communications}\ }\textbf {\bibinfo {volume}
  {185}},\ \bibinfo {pages} {2309} (\bibinfo {year} {2014})}\BibitemShut
  {NoStop}%
\bibitem [{\citenamefont {Marzari}\ \emph {et~al.}(2012)\citenamefont
  {Marzari}, \citenamefont {Mostofi}, \citenamefont {Yates}, \citenamefont
  {Souza},\ and\ \citenamefont {Vanderbilt}}]{wannierTheory}%
  \BibitemOpen
  \bibfield  {author} {\bibinfo {author} {\bibfnamefont {N.}~\bibnamefont
  {Marzari}}, \bibinfo {author} {\bibfnamefont {A.~A.}\ \bibnamefont
  {Mostofi}}, \bibinfo {author} {\bibfnamefont {J.~R.}\ \bibnamefont {Yates}},
  \bibinfo {author} {\bibfnamefont {I.}~\bibnamefont {Souza}}, \ and\ \bibinfo
  {author} {\bibfnamefont {D.}~\bibnamefont {Vanderbilt}},\ }\href {\doibase
  10.1103/RevModPhys.84.1419} {\bibfield  {journal} {\bibinfo  {journal} {Rev.
  Mod. Phys.}\ }\textbf {\bibinfo {volume} {84}},\ \bibinfo {pages} {1419}
  (\bibinfo {year} {2012})}\BibitemShut {NoStop}%
\bibitem [{\citenamefont {Li}\ \emph {et~al.}(2012)\citenamefont {Li},
  \citenamefont {Lu}, \citenamefont {Zhu}, \citenamefont {Ling}, \citenamefont
  {Qu},\ and\ \citenamefont {Sun}}]{Li2012}%
  \BibitemOpen
  \bibfield  {author} {\bibinfo {author} {\bibfnamefont {L.}~\bibnamefont
  {Li}}, \bibinfo {author} {\bibfnamefont {W.}~\bibnamefont {Lu}}, \bibinfo
  {author} {\bibfnamefont {X.}~\bibnamefont {Zhu}}, \bibinfo {author}
  {\bibfnamefont {L.}~\bibnamefont {Ling}}, \bibinfo {author} {\bibfnamefont
  {Z.}~\bibnamefont {Qu}}, \ and\ \bibinfo {author} {\bibfnamefont
  {Y.}~\bibnamefont {Sun}},\ }\href@noop {} {\bibfield  {journal} {\bibinfo
  {journal} {Europhysics Letters}\ }\textbf {\bibinfo {volume} {97}},\ \bibinfo
  {pages} {67005} (\bibinfo {year} {2012})}\BibitemShut {NoStop}%
\bibitem [{\citenamefont {Luo}\ \emph {et~al.}(2021)\citenamefont {Luo},
  \citenamefont {Obeysekera}, \citenamefont {Won}, \citenamefont {Sung},
  \citenamefont {Schnitzer}, \citenamefont {Hovden}, \citenamefont {Cheong},
  \citenamefont {Yang}, \citenamefont {Sun},\ and\ \citenamefont
  {Zhao}}]{luo2021ultrafast}%
  \BibitemOpen
  \bibfield  {author} {\bibinfo {author} {\bibfnamefont {X.}~\bibnamefont
  {Luo}}, \bibinfo {author} {\bibfnamefont {D.}~\bibnamefont {Obeysekera}},
  \bibinfo {author} {\bibfnamefont {C.}~\bibnamefont {Won}}, \bibinfo {author}
  {\bibfnamefont {S.~H.}\ \bibnamefont {Sung}}, \bibinfo {author}
  {\bibfnamefont {N.}~\bibnamefont {Schnitzer}}, \bibinfo {author}
  {\bibfnamefont {R.}~\bibnamefont {Hovden}}, \bibinfo {author} {\bibfnamefont
  {S.-W.}\ \bibnamefont {Cheong}}, \bibinfo {author} {\bibfnamefont
  {J.}~\bibnamefont {Yang}}, \bibinfo {author} {\bibfnamefont {K.}~\bibnamefont
  {Sun}}, \ and\ \bibinfo {author} {\bibfnamefont {L.}~\bibnamefont {Zhao}},\
  }\href@noop {} {\bibfield  {journal} {\bibinfo  {journal} {Physical review
  letters}\ }\textbf {\bibinfo {volume} {127}},\ \bibinfo {pages} {126401}
  (\bibinfo {year} {2021})}\BibitemShut {NoStop}%
\bibitem [{\citenamefont {Liu}\ \emph {et~al.}(2023)\citenamefont {Liu},
  \citenamefont {Qiu}, \citenamefont {He}, \citenamefont {Liu}, \citenamefont
  {Lin}, \citenamefont {Ma}, \citenamefont {Huang}, \citenamefont {Tang},
  \citenamefont {Xu}, \citenamefont {Watanabe} \emph
  {et~al.}}]{liu2023electrical}%
  \BibitemOpen
  \bibfield  {author} {\bibinfo {author} {\bibfnamefont {G.}~\bibnamefont
  {Liu}}, \bibinfo {author} {\bibfnamefont {T.}~\bibnamefont {Qiu}}, \bibinfo
  {author} {\bibfnamefont {K.}~\bibnamefont {He}}, \bibinfo {author}
  {\bibfnamefont {Y.}~\bibnamefont {Liu}}, \bibinfo {author} {\bibfnamefont
  {D.}~\bibnamefont {Lin}}, \bibinfo {author} {\bibfnamefont {Z.}~\bibnamefont
  {Ma}}, \bibinfo {author} {\bibfnamefont {Z.}~\bibnamefont {Huang}}, \bibinfo
  {author} {\bibfnamefont {W.}~\bibnamefont {Tang}}, \bibinfo {author}
  {\bibfnamefont {J.}~\bibnamefont {Xu}}, \bibinfo {author} {\bibfnamefont
  {K.}~\bibnamefont {Watanabe}},  \emph {et~al.},\ }\href@noop {} {\bibfield
  {journal} {\bibinfo  {journal} {Nature nanotechnology}\ }\textbf {\bibinfo
  {volume} {18}},\ \bibinfo {pages} {854} (\bibinfo {year} {2023})}\BibitemShut
  {NoStop}%
\bibitem [{\citenamefont {Kresse}\ and\ \citenamefont
  {Furthm\"uller}(1996)}]{VASP}%
  \BibitemOpen
  \bibfield  {author} {\bibinfo {author} {\bibfnamefont {G.}~\bibnamefont
  {Kresse}}\ and\ \bibinfo {author} {\bibfnamefont {J.}~\bibnamefont
  {Furthm\"uller}},\ }\href {\doibase 10.1103/PhysRevB.54.11169} {\bibfield
  {journal} {\bibinfo  {journal} {Phys. Rev. B}\ }\textbf {\bibinfo {volume}
  {54}},\ \bibinfo {pages} {11169} (\bibinfo {year} {1996})}\BibitemShut
  {NoStop}%
\bibitem [{\citenamefont {Kresse}\ and\ \citenamefont
  {Joubert}(1999)}]{VASP-PAW}%
  \BibitemOpen
  \bibfield  {author} {\bibinfo {author} {\bibfnamefont {G.}~\bibnamefont
  {Kresse}}\ and\ \bibinfo {author} {\bibfnamefont {D.}~\bibnamefont
  {Joubert}},\ }\href {\doibase 10.1103/PhysRevB.59.1758} {\bibfield  {journal}
  {\bibinfo  {journal} {Phys. Rev. B}\ }\textbf {\bibinfo {volume} {59}},\
  \bibinfo {pages} {1758} (\bibinfo {year} {1999})}\BibitemShut {NoStop}%
\bibitem [{\citenamefont {Perdew}\ \emph {et~al.}(2008)\citenamefont {Perdew},
  \citenamefont {Ruzsinszky}, \citenamefont {Csonka}, \citenamefont {Vydrov},
  \citenamefont {Scuseria}, \citenamefont {Constantin}, \citenamefont {Zhou},\
  and\ \citenamefont {Burke}}]{PBEsol}%
  \BibitemOpen
  \bibfield  {author} {\bibinfo {author} {\bibfnamefont {J.~P.}\ \bibnamefont
  {Perdew}}, \bibinfo {author} {\bibfnamefont {A.}~\bibnamefont {Ruzsinszky}},
  \bibinfo {author} {\bibfnamefont {G.~I.}\ \bibnamefont {Csonka}}, \bibinfo
  {author} {\bibfnamefont {O.~A.}\ \bibnamefont {Vydrov}}, \bibinfo {author}
  {\bibfnamefont {G.~E.}\ \bibnamefont {Scuseria}}, \bibinfo {author}
  {\bibfnamefont {L.~A.}\ \bibnamefont {Constantin}}, \bibinfo {author}
  {\bibfnamefont {X.}~\bibnamefont {Zhou}}, \ and\ \bibinfo {author}
  {\bibfnamefont {K.}~\bibnamefont {Burke}},\ }\href {\doibase
  10.1103/PhysRevLett.100.136406} {\bibfield  {journal} {\bibinfo  {journal}
  {Phys. Rev. Lett.}\ }\textbf {\bibinfo {volume} {100}},\ \bibinfo {pages}
  {136406} (\bibinfo {year} {2008})}\BibitemShut {NoStop}%
\bibitem [{\citenamefont {Bruneval}\ \emph {et~al.}(2014)\citenamefont
  {Bruneval}, \citenamefont {Crocombette}, \citenamefont {Gonze}, \citenamefont
  {Dorado}, \citenamefont {Torrent},\ and\ \citenamefont
  {Jollet}}]{Brubeval201}%
  \BibitemOpen
  \bibfield  {author} {\bibinfo {author} {\bibfnamefont {F.}~\bibnamefont
  {Bruneval}}, \bibinfo {author} {\bibfnamefont {J.-P.}\ \bibnamefont
  {Crocombette}}, \bibinfo {author} {\bibfnamefont {X.}~\bibnamefont {Gonze}},
  \bibinfo {author} {\bibfnamefont {B.}~\bibnamefont {Dorado}}, \bibinfo
  {author} {\bibfnamefont {M.}~\bibnamefont {Torrent}}, \ and\ \bibinfo
  {author} {\bibfnamefont {F.}~\bibnamefont {Jollet}},\ }\href {\doibase
  10.1103/PhysRevB.89.045116} {\bibfield  {journal} {\bibinfo  {journal} {Phys.
  Rev. B}\ }\textbf {\bibinfo {volume} {89}},\ \bibinfo {pages} {045116}
  (\bibinfo {year} {2014})}\BibitemShut {NoStop}%
\bibitem [{\citenamefont {Aroyo}\ \emph {et~al.}(2006)\citenamefont {Aroyo},
  \citenamefont {Kirov}, \citenamefont {Capillas}, \citenamefont {Perez-Mato},\
  and\ \citenamefont {Wondratschek}}]{Aroyo2006}%
  \BibitemOpen
  \bibfield  {author} {\bibinfo {author} {\bibfnamefont {M.~I.}\ \bibnamefont
  {Aroyo}}, \bibinfo {author} {\bibfnamefont {A.}~\bibnamefont {Kirov}},
  \bibinfo {author} {\bibfnamefont {C.}~\bibnamefont {Capillas}}, \bibinfo
  {author} {\bibfnamefont {J.}~\bibnamefont {Perez-Mato}}, \ and\ \bibinfo
  {author} {\bibfnamefont {H.}~\bibnamefont {Wondratschek}},\ }\href@noop {}
  {\bibfield  {journal} {\bibinfo  {journal} {Acta Crystallographica Section A:
  Foundations of Crystallography}\ }\textbf {\bibinfo {volume} {62}},\ \bibinfo
  {pages} {115} (\bibinfo {year} {2006})}\BibitemShut {NoStop}%
\bibitem [{\citenamefont {Wilson}\ \emph {et~al.}(1975)\citenamefont {Wilson},
  \citenamefont {Di~Salvo},\ and\ \citenamefont {Mahajan}}]{wilson1975charge}%
  \BibitemOpen
  \bibfield  {author} {\bibinfo {author} {\bibfnamefont {J.~A.}\ \bibnamefont
  {Wilson}}, \bibinfo {author} {\bibfnamefont {F.}~\bibnamefont {Di~Salvo}}, \
  and\ \bibinfo {author} {\bibfnamefont {S.}~\bibnamefont {Mahajan}},\
  }\href@noop {} {\bibfield  {journal} {\bibinfo  {journal} {Advances in
  Physics}\ }\textbf {\bibinfo {volume} {24}},\ \bibinfo {pages} {117}
  (\bibinfo {year} {1975})}\BibitemShut {NoStop}%
\bibitem [{\citenamefont {Hatch}\ and\ \citenamefont
  {Stokes}(2003)}]{Hatch2003}%
  \BibitemOpen
  \bibfield  {author} {\bibinfo {author} {\bibfnamefont {D.~M.}\ \bibnamefont
  {Hatch}}\ and\ \bibinfo {author} {\bibfnamefont {H.~T.}\ \bibnamefont
  {Stokes}},\ }\href {\doibase 10.1107/s0021889803005946} {\bibfield  {journal}
  {\bibinfo  {journal} {Journal of Applied Crystallography}\ }\textbf {\bibinfo
  {volume} {36}},\ \bibinfo {pages} {951} (\bibinfo {year} {2003})}\BibitemShut
  {NoStop}%
\bibitem [{\citenamefont {Benedek}\ \emph {et~al.}(2012)\citenamefont
  {Benedek}, \citenamefont {Mulder},\ and\ \citenamefont
  {Fennie}}]{Benedek2012}%
  \BibitemOpen
  \bibfield  {author} {\bibinfo {author} {\bibfnamefont {N.~A.}\ \bibnamefont
  {Benedek}}, \bibinfo {author} {\bibfnamefont {A.~T.}\ \bibnamefont {Mulder}},
  \ and\ \bibinfo {author} {\bibfnamefont {C.~J.}\ \bibnamefont {Fennie}},\
  }\href {\doibase 10.1016/j.jssc.2012.04.012} {\bibfield  {journal} {\bibinfo
  {journal} {Journal of Solid State Chemistry}\ }\textbf {\bibinfo {volume}
  {195}},\ \bibinfo {pages} {11} (\bibinfo {year} {2012})}\BibitemShut
  {NoStop}%
\bibitem [{\citenamefont {Benedek}\ and\ \citenamefont
  {Hayward}(2022)}]{Benedek2022}%
  \BibitemOpen
  \bibfield  {author} {\bibinfo {author} {\bibfnamefont {N.~A.}\ \bibnamefont
  {Benedek}}\ and\ \bibinfo {author} {\bibfnamefont {M.~A.}\ \bibnamefont
  {Hayward}},\ }\href {\doibase 10.1146/annurev-matsci-080819-010313}
  {\bibfield  {journal} {\bibinfo  {journal} {Annual Review of Materials
  Research}\ }\textbf {\bibinfo {volume} {52}},\ \bibinfo {pages} {331}
  (\bibinfo {year} {2022})}\BibitemShut {NoStop}%
\bibitem [{\citenamefont {Li}\ and\ \citenamefont {Birol}(2020)}]{Li2020}%
  \BibitemOpen
  \bibfield  {author} {\bibinfo {author} {\bibfnamefont {S.}~\bibnamefont
  {Li}}\ and\ \bibinfo {author} {\bibfnamefont {T.}~\bibnamefont {Birol}},\
  }\href {\doibase 10.1038/s41524-020-00436-x} {\bibfield  {journal} {\bibinfo
  {journal} {npj Computational Materials}\ }\textbf {\bibinfo {volume} {6}},\
  \bibinfo {pages} {168} (\bibinfo {year} {2020})}\BibitemShut {NoStop}%
\bibitem [{\citenamefont {Christensen}\ \emph {et~al.}(2021)\citenamefont
  {Christensen}, \citenamefont {Birol}, \citenamefont {Andersen},\ and\
  \citenamefont {Fernandes}}]{Christensen2021}%
  \BibitemOpen
  \bibfield  {author} {\bibinfo {author} {\bibfnamefont {M.~H.}\ \bibnamefont
  {Christensen}}, \bibinfo {author} {\bibfnamefont {T.}~\bibnamefont {Birol}},
  \bibinfo {author} {\bibfnamefont {B.~M.}\ \bibnamefont {Andersen}}, \ and\
  \bibinfo {author} {\bibfnamefont {R.~M.}\ \bibnamefont {Fernandes}},\ }\href
  {\doibase 10.1103/PhysRevB.104.214513} {\bibfield  {journal} {\bibinfo
  {journal} {Phys. Rev. B}\ }\textbf {\bibinfo {volume} {104}},\ \bibinfo
  {pages} {214513} (\bibinfo {year} {2021})}\BibitemShut {NoStop}%
\bibitem [{\citenamefont {Christensen}\ \emph {et~al.}(2022)\citenamefont
  {Christensen}, \citenamefont {Birol}, \citenamefont {Andersen},\ and\
  \citenamefont {Fernandes}}]{Christensen2022}%
  \BibitemOpen
  \bibfield  {author} {\bibinfo {author} {\bibfnamefont {M.~H.}\ \bibnamefont
  {Christensen}}, \bibinfo {author} {\bibfnamefont {T.}~\bibnamefont {Birol}},
  \bibinfo {author} {\bibfnamefont {B.~M.}\ \bibnamefont {Andersen}}, \ and\
  \bibinfo {author} {\bibfnamefont {R.~M.}\ \bibnamefont {Fernandes}},\ }\href
  {\doibase 10.1103/PhysRevB.106.144504} {\bibfield  {journal} {\bibinfo
  {journal} {Phys. Rev. B}\ }\textbf {\bibinfo {volume} {106}},\ \bibinfo
  {pages} {144504} (\bibinfo {year} {2022})}\BibitemShut {NoStop}%
\end{thebibliography}

\begin{thebibliography}{19}%
\makeatletter
\providecommand \@ifxundefined [1]{%
 \@ifx{#1\undefined}
}%
\providecommand \@ifnum [1]{%
 \ifnum #1\expandafter \@firstoftwo
 \else \expandafter \@secondoftwo
 \fi
}%
\providecommand \@ifx [1]{%
 \ifx #1\expandafter \@firstoftwo
 \else \expandafter \@secondoftwo
 \fi
}%
\providecommand \natexlab [1]{#1}%
\providecommand \enquote  [1]{``#1''}%
\providecommand \bibnamefont  [1]{#1}%
\providecommand \bibfnamefont [1]{#1}%
\providecommand \citenamefont [1]{#1}%
\providecommand \href@noop [0]{\@secondoftwo}%
\providecommand \href [0]{\begingroup \@sanitize@url \@href}%
\providecommand \@href[1]{\@@startlink{#1}\@@href}%
\providecommand \@@href[1]{\endgroup#1\@@endlink}%
\providecommand \@sanitize@url [0]{\catcode `\\12\catcode `\$12\catcode
  `\&12\catcode `\#12\catcode `\^12\catcode `\_12\catcode `\%12\relax}%
\providecommand \@@startlink[1]{}%
\providecommand \@@endlink[0]{}%
\providecommand \url  [0]{\begingroup\@sanitize@url \@url }%
\providecommand \@url [1]{\endgroup\@href {#1}{\urlprefix }}%
\providecommand \urlprefix  [0]{URL }%
\providecommand \Eprint [0]{\href }%
\providecommand \doibase [0]{http://dx.doi.org/}%
\providecommand \selectlanguage [0]{\@gobble}%
\providecommand \bibinfo  [0]{\@secondoftwo}%
\providecommand \bibfield  [0]{\@secondoftwo}%
\providecommand \translation [1]{[#1]}%
\providecommand \BibitemOpen [0]{}%
\providecommand \bibitemStop [0]{}%
\providecommand \bibitemNoStop [0]{.\EOS\space}%
\providecommand \EOS [0]{\spacefactor3000\relax}%
\providecommand \BibitemShut  [1]{\csname bibitem#1\endcsname}%
\let\auto@bib@innerbib\@empty
\bibitem [{\citenamefont {Kresse}\ and\ \citenamefont
  {Furthm\"uller}(1996)}]{VASP}%
  \BibitemOpen
  \bibfield  {author} {\bibinfo {author} {\bibfnamefont {G.}~\bibnamefont
  {Kresse}}\ and\ \bibinfo {author} {\bibfnamefont {J.}~\bibnamefont
  {Furthm\"uller}},\ }\href {\doibase 10.1103/PhysRevB.54.11169} {\bibfield
  {journal} {\bibinfo  {journal} {Phys. Rev. B}\ }\textbf {\bibinfo {volume}
  {54}},\ \bibinfo {pages} {11169} (\bibinfo {year} {1996})}\BibitemShut
  {NoStop}%
\bibitem [{\citenamefont {Kresse}\ and\ \citenamefont
  {Joubert}(1999)}]{VASP-PAW}%
  \BibitemOpen
  \bibfield  {author} {\bibinfo {author} {\bibfnamefont {G.}~\bibnamefont
  {Kresse}}\ and\ \bibinfo {author} {\bibfnamefont {D.}~\bibnamefont
  {Joubert}},\ }\href {\doibase 10.1103/PhysRevB.59.1758} {\bibfield  {journal}
  {\bibinfo  {journal} {Phys. Rev. B}\ }\textbf {\bibinfo {volume} {59}},\
  \bibinfo {pages} {1758} (\bibinfo {year} {1999})}\BibitemShut {NoStop}%
\bibitem [{\citenamefont {Perdew}\ \emph {et~al.}(2008)\citenamefont {Perdew},
  \citenamefont {Ruzsinszky}, \citenamefont {Csonka}, \citenamefont {Vydrov},
  \citenamefont {Scuseria}, \citenamefont {Constantin}, \citenamefont {Zhou},\
  and\ \citenamefont {Burke}}]{PBEsol}%
  \BibitemOpen
  \bibfield  {author} {\bibinfo {author} {\bibfnamefont {J.~P.}\ \bibnamefont
  {Perdew}}, \bibinfo {author} {\bibfnamefont {A.}~\bibnamefont {Ruzsinszky}},
  \bibinfo {author} {\bibfnamefont {G.~I.}\ \bibnamefont {Csonka}}, \bibinfo
  {author} {\bibfnamefont {O.~A.}\ \bibnamefont {Vydrov}}, \bibinfo {author}
  {\bibfnamefont {G.~E.}\ \bibnamefont {Scuseria}}, \bibinfo {author}
  {\bibfnamefont {L.~A.}\ \bibnamefont {Constantin}}, \bibinfo {author}
  {\bibfnamefont {X.}~\bibnamefont {Zhou}}, \ and\ \bibinfo {author}
  {\bibfnamefont {K.}~\bibnamefont {Burke}},\ }\href {\doibase
  10.1103/PhysRevLett.100.136406} {\bibfield  {journal} {\bibinfo  {journal}
  {Phys. Rev. Lett.}\ }\textbf {\bibinfo {volume} {100}},\ \bibinfo {pages}
  {136406} (\bibinfo {year} {2008})}\BibitemShut {NoStop}%
\bibitem [{\citenamefont {Pizzi}\ \emph {et~al.}(2014)\citenamefont {Pizzi},
  \citenamefont {Volja}, \citenamefont {Kozinsky}, \citenamefont {Fornari},\
  and\ \citenamefont {Marzari}}]{pizzi2014boltzwann}%
  \BibitemOpen
  \bibfield  {author} {\bibinfo {author} {\bibfnamefont {G.}~\bibnamefont
  {Pizzi}}, \bibinfo {author} {\bibfnamefont {D.}~\bibnamefont {Volja}},
  \bibinfo {author} {\bibfnamefont {B.}~\bibnamefont {Kozinsky}}, \bibinfo
  {author} {\bibfnamefont {M.}~\bibnamefont {Fornari}}, \ and\ \bibinfo
  {author} {\bibfnamefont {N.}~\bibnamefont {Marzari}},\ }\href@noop {}
  {\bibfield  {journal} {\bibinfo  {journal} {Computer Physics Communications}\
  }\textbf {\bibinfo {volume} {185}},\ \bibinfo {pages} {422} (\bibinfo {year}
  {2014})}\BibitemShut {NoStop}%
\bibitem [{\citenamefont {Mostofi}\ \emph {et~al.}(2014)\citenamefont
  {Mostofi}, \citenamefont {Yates}, \citenamefont {Pizzi}, \citenamefont {Lee},
  \citenamefont {Souza}, \citenamefont {Vanderbilt},\ and\ \citenamefont
  {Marzari}}]{wannier90}%
  \BibitemOpen
  \bibfield  {author} {\bibinfo {author} {\bibfnamefont {A.~A.}\ \bibnamefont
  {Mostofi}}, \bibinfo {author} {\bibfnamefont {J.~R.}\ \bibnamefont {Yates}},
  \bibinfo {author} {\bibfnamefont {G.}~\bibnamefont {Pizzi}}, \bibinfo
  {author} {\bibfnamefont {Y.-S.}\ \bibnamefont {Lee}}, \bibinfo {author}
  {\bibfnamefont {I.}~\bibnamefont {Souza}}, \bibinfo {author} {\bibfnamefont
  {D.}~\bibnamefont {Vanderbilt}}, \ and\ \bibinfo {author} {\bibfnamefont
  {N.}~\bibnamefont {Marzari}},\ }\href@noop {} {\bibfield  {journal} {\bibinfo
   {journal} {Computer Physics Communications}\ }\textbf {\bibinfo {volume}
  {185}},\ \bibinfo {pages} {2309} (\bibinfo {year} {2014})}\BibitemShut
  {NoStop}%
\bibitem [{\citenamefont {Bruneval}\ \emph {et~al.}(2014)\citenamefont
  {Bruneval}, \citenamefont {Crocombette}, \citenamefont {Gonze}, \citenamefont
  {Dorado}, \citenamefont {Torrent},\ and\ \citenamefont
  {Jollet}}]{Brubeval201}%
  \BibitemOpen
  \bibfield  {author} {\bibinfo {author} {\bibfnamefont {F.}~\bibnamefont
  {Bruneval}}, \bibinfo {author} {\bibfnamefont {J.-P.}\ \bibnamefont
  {Crocombette}}, \bibinfo {author} {\bibfnamefont {X.}~\bibnamefont {Gonze}},
  \bibinfo {author} {\bibfnamefont {B.}~\bibnamefont {Dorado}}, \bibinfo
  {author} {\bibfnamefont {M.}~\bibnamefont {Torrent}}, \ and\ \bibinfo
  {author} {\bibfnamefont {F.}~\bibnamefont {Jollet}},\ }\href {\doibase
  10.1103/PhysRevB.89.045116} {\bibfield  {journal} {\bibinfo  {journal} {Phys.
  Rev. B}\ }\textbf {\bibinfo {volume} {89}},\ \bibinfo {pages} {045116}
  (\bibinfo {year} {2014})}\BibitemShut {NoStop}%
\bibitem [{\citenamefont {Dresselhaus}\ \emph {et~al.}(2008)\citenamefont
  {Dresselhaus}, \citenamefont {Dresselhaus},\ and\ \citenamefont
  {Jorio}}]{dresselhaus}%
  \BibitemOpen
  \bibfield  {author} {\bibinfo {author} {\bibfnamefont {M.~S.}\ \bibnamefont
  {Dresselhaus}}, \bibinfo {author} {\bibfnamefont {G.}~\bibnamefont
  {Dresselhaus}}, \ and\ \bibinfo {author} {\bibfnamefont {A.}~\bibnamefont
  {Jorio}},\ }\href@noop {} {\emph {\bibinfo {title} {Applications of group
  theory to the physics of solids}}}\ (\bibinfo  {publisher} {Springer
  Berlin},\ \bibinfo {year} {2008})\BibitemShut {NoStop}%
\bibitem [{\citenamefont {Nye}(1985)}]{Nye1985book}%
  \BibitemOpen
  \bibfield  {author} {\bibinfo {author} {\bibfnamefont {J.~F.}\ \bibnamefont
  {Nye}},\ }\href@noop {} {\emph {\bibinfo {title} {Physical properties of
  crystals: their representation by tensors and matrices}}}\ (\bibinfo
  {publisher} {Oxford university press},\ \bibinfo {year} {1985})\BibitemShut
  {NoStop}%
\bibitem [{\citenamefont {Aroyo}\ \emph {et~al.}(2006)\citenamefont {Aroyo},
  \citenamefont {Kirov}, \citenamefont {Capillas}, \citenamefont {Perez-Mato},\
  and\ \citenamefont {Wondratschek}}]{Aroyo2006}%
  \BibitemOpen
  \bibfield  {author} {\bibinfo {author} {\bibfnamefont {M.~I.}\ \bibnamefont
  {Aroyo}}, \bibinfo {author} {\bibfnamefont {A.}~\bibnamefont {Kirov}},
  \bibinfo {author} {\bibfnamefont {C.}~\bibnamefont {Capillas}}, \bibinfo
  {author} {\bibfnamefont {J.}~\bibnamefont {Perez-Mato}}, \ and\ \bibinfo
  {author} {\bibfnamefont {H.}~\bibnamefont {Wondratschek}},\ }\href@noop {}
  {\bibfield  {journal} {\bibinfo  {journal} {Acta Crystallographica Section A:
  Foundations of Crystallography}\ }\textbf {\bibinfo {volume} {62}},\ \bibinfo
  {pages} {115} (\bibinfo {year} {2006})}\BibitemShut {NoStop}%
\bibitem [{\citenamefont {Wilson}\ \emph {et~al.}(1975)\citenamefont {Wilson},
  \citenamefont {Di~Salvo},\ and\ \citenamefont {Mahajan}}]{wilson1975charge}%
  \BibitemOpen
  \bibfield  {author} {\bibinfo {author} {\bibfnamefont {J.~A.}\ \bibnamefont
  {Wilson}}, \bibinfo {author} {\bibfnamefont {F.}~\bibnamefont {Di~Salvo}}, \
  and\ \bibinfo {author} {\bibfnamefont {S.}~\bibnamefont {Mahajan}},\
  }\href@noop {} {\bibfield  {journal} {\bibinfo  {journal} {Advances in
  Physics}\ }\textbf {\bibinfo {volume} {24}},\ \bibinfo {pages} {117}
  (\bibinfo {year} {1975})}\BibitemShut {NoStop}%
\bibitem [{\citenamefont {Liu}\ \emph {et~al.}(2023)\citenamefont {Liu},
  \citenamefont {Qiu}, \citenamefont {He}, \citenamefont {Liu}, \citenamefont
  {Lin}, \citenamefont {Ma}, \citenamefont {Huang}, \citenamefont {Tang},
  \citenamefont {Xu}, \citenamefont {Watanabe} \emph
  {et~al.}}]{liu2023electrical}%
  \BibitemOpen
  \bibfield  {author} {\bibinfo {author} {\bibfnamefont {G.}~\bibnamefont
  {Liu}}, \bibinfo {author} {\bibfnamefont {T.}~\bibnamefont {Qiu}}, \bibinfo
  {author} {\bibfnamefont {K.}~\bibnamefont {He}}, \bibinfo {author}
  {\bibfnamefont {Y.}~\bibnamefont {Liu}}, \bibinfo {author} {\bibfnamefont
  {D.}~\bibnamefont {Lin}}, \bibinfo {author} {\bibfnamefont {Z.}~\bibnamefont
  {Ma}}, \bibinfo {author} {\bibfnamefont {Z.}~\bibnamefont {Huang}}, \bibinfo
  {author} {\bibfnamefont {W.}~\bibnamefont {Tang}}, \bibinfo {author}
  {\bibfnamefont {J.}~\bibnamefont {Xu}}, \bibinfo {author} {\bibfnamefont
  {K.}~\bibnamefont {Watanabe}},  \emph {et~al.},\ }\href@noop {} {\bibfield
  {journal} {\bibinfo  {journal} {Nature nanotechnology}\ }\textbf {\bibinfo
  {volume} {18}},\ \bibinfo {pages} {854} (\bibinfo {year} {2023})}\BibitemShut
  {NoStop}%
\bibitem [{\citenamefont {Luo}\ \emph {et~al.}(2021)\citenamefont {Luo},
  \citenamefont {Obeysekera}, \citenamefont {Won}, \citenamefont {Sung},
  \citenamefont {Schnitzer}, \citenamefont {Hovden}, \citenamefont {Cheong},
  \citenamefont {Yang}, \citenamefont {Sun},\ and\ \citenamefont
  {Zhao}}]{luo2021ultrafast}%
  \BibitemOpen
  \bibfield  {author} {\bibinfo {author} {\bibfnamefont {X.}~\bibnamefont
  {Luo}}, \bibinfo {author} {\bibfnamefont {D.}~\bibnamefont {Obeysekera}},
  \bibinfo {author} {\bibfnamefont {C.}~\bibnamefont {Won}}, \bibinfo {author}
  {\bibfnamefont {S.~H.}\ \bibnamefont {Sung}}, \bibinfo {author}
  {\bibfnamefont {N.}~\bibnamefont {Schnitzer}}, \bibinfo {author}
  {\bibfnamefont {R.}~\bibnamefont {Hovden}}, \bibinfo {author} {\bibfnamefont
  {S.-W.}\ \bibnamefont {Cheong}}, \bibinfo {author} {\bibfnamefont
  {J.}~\bibnamefont {Yang}}, \bibinfo {author} {\bibfnamefont {K.}~\bibnamefont
  {Sun}}, \ and\ \bibinfo {author} {\bibfnamefont {L.}~\bibnamefont {Zhao}},\
  }\href@noop {} {\bibfield  {journal} {\bibinfo  {journal} {Physical review
  letters}\ }\textbf {\bibinfo {volume} {127}},\ \bibinfo {pages} {126401}
  (\bibinfo {year} {2021})}\BibitemShut {NoStop}%
\bibitem [{\citenamefont {Li}\ \emph {et~al.}(2012)\citenamefont {Li},
  \citenamefont {Lu}, \citenamefont {Zhu}, \citenamefont {Ling}, \citenamefont
  {Qu},\ and\ \citenamefont {Sun}}]{Li2012}%
  \BibitemOpen
  \bibfield  {author} {\bibinfo {author} {\bibfnamefont {L.}~\bibnamefont
  {Li}}, \bibinfo {author} {\bibfnamefont {W.}~\bibnamefont {Lu}}, \bibinfo
  {author} {\bibfnamefont {X.}~\bibnamefont {Zhu}}, \bibinfo {author}
  {\bibfnamefont {L.}~\bibnamefont {Ling}}, \bibinfo {author} {\bibfnamefont
  {Z.}~\bibnamefont {Qu}}, \ and\ \bibinfo {author} {\bibfnamefont
  {Y.}~\bibnamefont {Sun}},\ }\href@noop {} {\bibfield  {journal} {\bibinfo
  {journal} {Europhysics Letters}\ }\textbf {\bibinfo {volume} {97}},\ \bibinfo
  {pages} {67005} (\bibinfo {year} {2012})}\BibitemShut {NoStop}%
\bibitem [{\citenamefont {Hatch}\ and\ \citenamefont
  {Stokes}(2003)}]{Hatch2003}%
  \BibitemOpen
  \bibfield  {author} {\bibinfo {author} {\bibfnamefont {D.~M.}\ \bibnamefont
  {Hatch}}\ and\ \bibinfo {author} {\bibfnamefont {H.~T.}\ \bibnamefont
  {Stokes}},\ }\href {\doibase 10.1107/s0021889803005946} {\bibfield  {journal}
  {\bibinfo  {journal} {Journal of Applied Crystallography}\ }\textbf {\bibinfo
  {volume} {36}},\ \bibinfo {pages} {951} (\bibinfo {year} {2003})}\BibitemShut
  {NoStop}%
\bibitem [{\citenamefont {Benedek}\ \emph {et~al.}(2012)\citenamefont
  {Benedek}, \citenamefont {Mulder},\ and\ \citenamefont
  {Fennie}}]{Benedek2012}%
  \BibitemOpen
  \bibfield  {author} {\bibinfo {author} {\bibfnamefont {N.~A.}\ \bibnamefont
  {Benedek}}, \bibinfo {author} {\bibfnamefont {A.~T.}\ \bibnamefont {Mulder}},
  \ and\ \bibinfo {author} {\bibfnamefont {C.~J.}\ \bibnamefont {Fennie}},\
  }\href {\doibase 10.1016/j.jssc.2012.04.012} {\bibfield  {journal} {\bibinfo
  {journal} {Journal of Solid State Chemistry}\ }\textbf {\bibinfo {volume}
  {195}},\ \bibinfo {pages} {11} (\bibinfo {year} {2012})}\BibitemShut
  {NoStop}%
\bibitem [{\citenamefont {Benedek}\ and\ \citenamefont
  {Hayward}(2022)}]{Benedek2022}%
  \BibitemOpen
  \bibfield  {author} {\bibinfo {author} {\bibfnamefont {N.~A.}\ \bibnamefont
  {Benedek}}\ and\ \bibinfo {author} {\bibfnamefont {M.~A.}\ \bibnamefont
  {Hayward}},\ }\href {\doibase 10.1146/annurev-matsci-080819-010313}
  {\bibfield  {journal} {\bibinfo  {journal} {Annual Review of Materials
  Research}\ }\textbf {\bibinfo {volume} {52}},\ \bibinfo {pages} {331}
  (\bibinfo {year} {2022})}\BibitemShut {NoStop}%
\bibitem [{\citenamefont {Li}\ and\ \citenamefont {Birol}(2020)}]{Li2020}%
  \BibitemOpen
  \bibfield  {author} {\bibinfo {author} {\bibfnamefont {S.}~\bibnamefont
  {Li}}\ and\ \bibinfo {author} {\bibfnamefont {T.}~\bibnamefont {Birol}},\
  }\href {\doibase 10.1038/s41524-020-00436-x} {\bibfield  {journal} {\bibinfo
  {journal} {npj Computational Materials}\ }\textbf {\bibinfo {volume} {6}},\
  \bibinfo {pages} {168} (\bibinfo {year} {2020})}\BibitemShut {NoStop}%
\bibitem [{\citenamefont {Christensen}\ \emph {et~al.}(2021)\citenamefont
  {Christensen}, \citenamefont {Birol}, \citenamefont {Andersen},\ and\
  \citenamefont {Fernandes}}]{Christensen2021}%
  \BibitemOpen
  \bibfield  {author} {\bibinfo {author} {\bibfnamefont {M.~H.}\ \bibnamefont
  {Christensen}}, \bibinfo {author} {\bibfnamefont {T.}~\bibnamefont {Birol}},
  \bibinfo {author} {\bibfnamefont {B.~M.}\ \bibnamefont {Andersen}}, \ and\
  \bibinfo {author} {\bibfnamefont {R.~M.}\ \bibnamefont {Fernandes}},\ }\href
  {\doibase 10.1103/PhysRevB.104.214513} {\bibfield  {journal} {\bibinfo
  {journal} {Phys. Rev. B}\ }\textbf {\bibinfo {volume} {104}},\ \bibinfo
  {pages} {214513} (\bibinfo {year} {2021})}\BibitemShut {NoStop}%
\bibitem [{\citenamefont {Christensen}\ \emph {et~al.}(2022)\citenamefont
  {Christensen}, \citenamefont {Birol}, \citenamefont {Andersen},\ and\
  \citenamefont {Fernandes}}]{Christensen2022}%
  \BibitemOpen
  \bibfield  {author} {\bibinfo {author} {\bibfnamefont {M.~H.}\ \bibnamefont
  {Christensen}}, \bibinfo {author} {\bibfnamefont {T.}~\bibnamefont {Birol}},
  \bibinfo {author} {\bibfnamefont {B.~M.}\ \bibnamefont {Andersen}}, \ and\
  \bibinfo {author} {\bibfnamefont {R.~M.}\ \bibnamefont {Fernandes}},\ }\href
  {\doibase 10.1103/PhysRevB.106.144504} {\bibfield  {journal} {\bibinfo
  {journal} {Phys. Rev. B}\ }\textbf {\bibinfo {volume} {106}},\ \bibinfo
  {pages} {144504} (\bibinfo {year} {2022})}\BibitemShut {NoStop}%
\end{thebibliography}
\end{document}